\documentclass[11pt]{article}

\usepackage[left=2cm, right=2cm, top=2.5cm, bottom=2.5cm]{geometry}
\usepackage[x11names]{xcolor}
\usepackage{fancyhdr, amssymb, cancel, amsmath, graphicx, pgfplots, tikz}

\newcommand{\stylecolor}{black}

\usepackage[colorlinks=true, urlcolor=violet, linkcolor=blue, citecolor=red, hyperindex=true, linktocpage=true]{hyperref}

\usepackage[explicit]{titlesec}

\newcommand*\sectionlabel{}
\titleformat{\section}
  {\gdef\sectionlabel{}
   \Large\bfseries\scshape}
  {\gdef\sectionlabel{\thesection. }}{0pt}
  {\begin{tikzpicture}[remember picture,overlay]
	\draw (-0.2, 0) node[right] {\textsf{\sectionlabel#1}};
	\draw[thick] (0, -0.4) -- (\textwidth, -0.4);
       \end{tikzpicture}
  }
\titlespacing*{\section}{0pt}{15pt}{20pt}

\newcommand*\subsectionlabel{}
\titleformat{\subsection}
  {\gdef\subsectionlabel{}
   \large\bfseries\scshape}
  {\gdef\subsectionlabel{\thesubsection.\ \  }}{0pt}
  {\begin{tikzpicture}[remember picture,overlay]
    	\draw (-0.15, 0) node[right] {\textsf{\subsectionlabel#1}};
       \end{tikzpicture}
  }
\titlespacing*{\subsection}{0pt}{10pt}{10pt}

\newcommand*\subsubsectionlabel{}
\titleformat{\subsubsection}
  {\gdef\subsubsectionlabel{}
   \bfseries\scshape}
  {\gdef\subsubsectionlabel{\thesubsubsection.\ \  }}{0pt}
  {\begin{tikzpicture}[remember picture,overlay]
    	\draw (-0.15, 0) node[right] {\textsf{\subsubsectionlabel#1}};
       \end{tikzpicture}
  }
\titlespacing*{\subsubsection}{0pt}{7pt}{7pt}

\pgfplotsset{every axis legend/.append style={at={(1.02,1)},anchor=north west}}

\newcommand{\titletext}{Phenomenology of non-relativistic parity-violating hydrodynamics in 2+1 dimensions}

\begin{document}

\pagestyle{fancy}
\renewcommand{\headrulewidth}{0pt}
\fancyhead{}

\fancyfoot{}
\fancyfoot[C] {\textsf{\textbf{\thepage}}}

\begin{equation*}
\begin{tikzpicture}
\draw (0.5\textwidth, -3) node[text width = \textwidth] {{\huge \begin{center} \color{\stylecolor} \textsf{\textbf{\titletext}} \end{center}}};
\end{tikzpicture}
\end{equation*}
\begin{equation*}
\begin{tikzpicture}
\draw (0.5\textwidth, 0.1) node[text width=\textwidth] {\large \color{black} \textsf{Andrew Lucas$^a$ and Piotr Sur\'owka$^{b,a}$}};
\draw (0.5\textwidth, -0.5) node[text width=\textwidth] {\small  $^a$\textsf{Department of Physics, Harvard University, Cambridge, MA 02138, USA}};
\draw (0.5\textwidth, -1) node[text width=\textwidth] {\small  $^b$\textsf{Center for the Fundamental Laws of Nature, Harvard University, Cambridge, MA 02138, USA}};
\end{tikzpicture}
\end{equation*}
\begin{equation*}
\begin{tikzpicture}
\draw (0.5\textwidth, -6) node[below, text width=0.8\textwidth] {\small Parity-violating fluids in two spatial dimensions can appear in a variety of contexts such as liquid crystal films, anyon fluids, and quantum Hall fluids.   Nonetheless, the consequences of parity-violation on the solutions to the equations of motion are largely unexplored.   In this paper, we explore phenomenological consequences of parity-violation through simple, illustrative examples.  Although incompressible velocity fields are essentially unchanged by parity violation, we discuss examples where parity violation plays a role at boundaries, or in the dynamics of temperature.   We then discuss new types of compressible flows which only exist in a parity-violating fluid, including new sound waves, and solitons in the dissipationless limit.   We conclude with a discussion of some curious features in Rayleigh-B\'enard convection of a parity-violating fluid.};
\end{tikzpicture}
\end{equation*}
\begin{equation*}
\begin{tikzpicture}
\draw (0, -13.1) node[right, text width=0.5\textwidth] {\texttt{lucas@fas.harvard.edu \\ surowka@physics.harvard.edu}};
\draw (\textwidth, -13.1) node[left] {\textsf{\today}};
\end{tikzpicture}
\end{equation*}

\tableofcontents

\section{Introduction}
Recently, theoretical physicists have renewed interest in hydrodynamics due to the rapid thermalization of strongly-coupled quantum field theories.   Strong evidence for this has emerged recently:  e.g., via the AdS/CFT correspondence \cite{chesler}, and through recent work detailing the rapid emergence of steady-state transport in higher dimensions \cite{bhaseen1, scooper}.    Hydrodynamics provides a consistent macroscopic description of the dynamics of strongly-coupled, many-body systems, at long time and length scales compared to the appropriate microscopic scales, such as the mean free time/path.   For arbitrary thermalizing systems in $d$ spatial dimensions, the equations of hydrodynamics are: \begin{equation}
\partial_\mu j^\mu_A=\partial_t j^t_A + \partial_i j^i_A = 0,
\end{equation}where $j^t_A$ is a density and $j^i_A$ is a current associated with conserved charges in the theory, which we label with index $A$.   The expressions for $j^\mu_A$ are organized in a perturbative gradient expansion as $j^\mu_{0A} + j^\mu_{1A}+\cdots$, where each term $j^\mu_{nA}$ has $n$ spatial derivatives.   One must formally determine $j^\mu_{nA}$ by a microscopic computation -- however, in many cases, the underlying symmetries of the theory, as well as the second law of thermodynamics, provide strong constraints \cite{kadanoff, landau}.   In most cases this gradient expansion is valid for theories with underlying dynamics which is either classical or quantum.   To put it simply, one should think of hydrodynamics as \emph{perturbation theory around thermodynamics}, where the perturbative parameters are the number of derivatives present in the currents $j^\mu_{nA}$.

The most familiar fluids in physics are normal liquids and gases like air and water.   In these fluids, quantum effects play no role over distances larger than the atomic and molecular scale, and for these gases one can directly imagine computing the viscosity from a classical computation using the Boltzmann equation.    On molecular scales, these fluids are very close to equilibrium, but since we can observe the motion of these fluids on orders of magnitude larger scales, we can observe flows where gradients in density or velocities are quite large, compared to our macroscopic length scale.

These are not the only fluids in nature.  A slightly more exotic example consists of a classical plasma of ionized gas, where electrons and ions are separated.  In this case, the charged nature of the microscopic constituents is often quite important \cite{landaukinetic}: this manifests itself in the addition of new terms in the fluid equations which couple the fluid to these external fields, though the general framework of hydrodynamics remains valid.    Another important example of a fluid is the electron plasma in a metal \cite{landaukinetic}.  Although microscopically this is a very different plasma consisting of mobile electrons sitting in a bath of positively-charged (and approximately stationary) ions, if we only ask questions about the transport of particles, energy and momentum on long distances compared to the mean free path of electrons, this plasma must also obey the same hydrodynamic equations as the classical plasma (though the particular values of, say, the pressure or viscosity, may differ).\footnote{Formally, the presence of a lattice breaks some of the spacetime symmetry by picking out a preferred rest frame for the electronic plasma, and may also break some of the rotational symmetry of the spacetime.  By simply reducing the symmetry constraints on the gradient expansion, hydrodynamics can easily account for these phenomena.}

Right now, the direct observation of hydrodynamic modes in a quantum system (without superfluidity) in a laboratory is very challenging.  By this, we mean that it is very difficult to observe the viscous dissipation of energy in a quantum fluid, as an example.  This is because there are effective ``friction" forces due to the lattice that can remove momentum from the electron fluid by exciting lattice vibrations.  Nonetheless, cold atomic gases are allowing for experimental realizations of the out-of -equilibrium dynamics of exotic, interacting quantum systems \cite{goldman} for which hydrodynamics is often an effective description, and we hope that these experiments will soon be able directly observe hydrodynamic modes in a strongly-coupled quantum theory.   For alternative ideas, see the recent work \cite{zaanen}.

\subsection{Parity-Violating Fluids}
One of the fascinating possibilities in hydrodynamics is the reduction of the usual symmetries of space and time that classical fluids like water possess.   Perhaps the most important example would be the breaking of parity symmetry, while preserving spatial isotropy (rotational invariance).   In 3+1 dimensions (by this we mean 3 spatial and 1 temporal dimension), parity acts by changing $(x,y,z) \rightarrow -(x,y,z)$.   As this linear transformation has determinant $-1$, it is \emph{not} equivalent to a rotation -- it is a discrete symmetry that essentially says that the spacetime does not pick out a preferred orientation.   Remarkably, this symmetry is often broken in nature.   An important example in the non-relativistic world is that many chemical compounds are \emph{chiral} -- if we perform this parity operation on the molecule, we end up with a different molecule!   Parity plays an important role in many biochemical processes, where only the right chirality molecule will ``fit" into a protein, as an example.   A fluid consisting of these chiral molecules would also break parity symmetry, and this would allow us to add more terms in the hydrodynamic gradient expansion.   In this paper, we will be interested in lower dimensional physics in 2+1 dimensions, where parity is a bit different.   Here, parity symmetry corresponds to \emph{either} $x\rightarrow -x$ \emph{or} $y\rightarrow -y$:  applying both is simply equivalent to a rotation.   A simple way to obtain a parity-violating fluid in (effectively) two dimensions  would be to take a film of a classical fluid of chiral molecules, but we now know of many more examples, as we will elaborate below.

It is important to stress that we are referring to a parity-violating fluid as a fluid where the microscopic components break parity.   It is of course easy to construct a flow which violates parity -- for example, placing a vortex in a two-dimensional liquid film breaks parity, because under parity the circulation of the vortex switches sign.   In parity-violating hydrodynamics, we will see that there are additional transport coefficients which are odd under parity -- this is a fundamentally different physical effect.

  In 3+1 dimensions it was shown that such terms can arise because of quantum anomalies of the underlying field theory \cite{surowkaanom}.   Associated transport was explained to be non-dissipative and the pertinent transport coefficients were constrained by the second law of thermodynamics or, more generally, from the relevant effective actions \cite{logan1, logan2, logan3, logan4, logan5}. In 2+1 dimensions anomalies do not exist, but similar arguments based on the entropy current and partition functions apply, and parity-odd contributions to the hydrodynamic gradient expansion were consistently derived \cite{jensen, cai, haehl, kaminski, sonwu, geracie}. One possible term that can be added involves the dissipationless Hall viscosity.  Hydrodynamic arguments, based on the second law of thermodynamics, tell us that the Hall viscosity may be an arbitrary function of entropy and particle density \cite{haehl}, in a generic theory.\footnote{However, for some specific theories such as quantum Hall states, the Hall viscosity is highly constrained \cite{rezayi}.}

Although relativistic fluids in any number of spatial dimensions $d>1$ exist at quantum critical points in metals,\footnote{See \cite{bhaseen1}  for a recent example of the consequences of hydrodynamics on two interacting quantum critical heat baths.} our intuition about the hydrodynamics of relativistic fluids is rather limited, especially in the context of condensed matter.   Our goal in this paper is to understand the phenomenology of parity-violating isotropic fluids, and to this end, we will take the non-relativistic limit of the equations, where we have more intuition for the possible types of dynamics.   We also hope to reach various communities for which hydrodynamics is a useful tool.  As many communities, such as condensed matter physics, require a systematic treatment of charge in the hydrodynamic expansion,  we will be using an unified approach which contains the fluid equations with a background electromagnetic field.  This allows to investigate parity-breaking effects in condensed matter systems and plasma physics.
In practice, for this paper one can consistently set the electric fields to zero; when the magnetic field is important to get a correction to the parity-even set-up, we emphasize that in the text.

Parity odd corrections to the non-relativistic hydrodynamic equations were recently derived in \cite{kaminski}.   They include the effect of Hall viscosity in the Navier-Stokes equations,  first explored in \cite{avron}.   The large number of new terms in the conservation laws suggests that parity-odd effects may be phenomenologically important for certain classes of flows.   For example an external probe sitting in an incompressible fluid has been studied, where there are new stresses normal to the surface \cite{lapa}.   However, the effects of parity-violation on the hydrodynamic flows themselves are poorly understood, as the Hall viscosity is effectively a ``topological" surface term in the incompressible Navier-Stokes equation, and so more terms must be added to see new physics away from boundaries.\footnote{This logic is only valid for fluid flows on the plane.  See \cite{wiegmann} for flows on more exotic spaces.}  Although the above developments are often quite formal and abstract, a detailed understanding of solutions to parity-violating hydrodynamics might shed new light on the dynamics of both quantum and classical parity-violating systems.  These can include quantum Hall states \cite{rezayi, avron2, haldane, read, choyos}, topological insulators \cite{hughes, barkeshli}, fluids of chiral molecules \cite{andreev}, liquid crystals \cite{forster, martin}, chiral superfluids \cite{hoyos} and anyon fluids \cite{wenzee}.  In this paper we will focus on the 2+1 dimensional flows and leave higher dimensional cases for future investigation.

The rest of this paper is organized as follows.   We conclude this introductory section with a summary of the most general parity-violating non-relativistic hydrodynamics to lowest order in the gradient expansion, and a definition and/or summary of common fluid mechanics shorthand we will use.   In Section \ref{sec2} we provide a simple example of a problem in heat flow where parity-violating boundary conditions can have dramatic effects.   In Section \ref{sec3} we discuss incompressible fluids.   In Section \ref{sec4} we discuss sound propagation.  In Section \ref{sec5} we discuss new types of dissipationless parity-violating fluid flows.   Section \ref{sec6} describes a parity-violating analogue of the Rayleigh-B\'enard convective instability.

\subsection{The Equations of Motion}
Relativistic physical theories have Lorentz invariance, and many non-relativistic sets of equations with Galilean invariance can be viewed as a limiting case of underlying Lorentz-invariant relativistic equations, in the limit where all fluid velocities are very small compared to the speed of light.  For example, viscous Navier-Stokes equations may be obtained from relativistic hydrodynamics \cite{fouxon,minwallaNR}.\footnote{Strictly speaking, these relativistic viscous theories are not causal.   This does not mean that they are not proper relativistic theories of hydrodynamics.  Hydrodynamics is a gradient expansion and therefore is only valid for slow dynamics on long length scales -- the modes in these theories which violate causality are outside of the regime of validity of hydrodynamics.}  A rigorous derivation of this non-relativistic limit for parity-violating fluids, based on non-relativistic coordinate invariance, was done in \cite{kaminski}, following earlier developments valid for parity preserving theories \cite{wingate}.  In this section we summarize the results of \cite{kaminski}.   There are some subtleties with the extraction of non-relativistic thermodynamic quantities from their relativistic counterparts; for the purposes of this paper, we will take the hydrodynamic expansion as given and explore its phenomenology.

For simplicity, we assume that we have a single-component fluid with particles of mass $m$ and charge $q$, in external electric field $E^i$ and magnetic field $B$.   As with any classical plasma, if $q\ne 0$, we are implicitly assuming that there is some oppositely charged heavy stuff that does not move, but cancels off the large electromagnetic fields from the fluid constituents.   In fact, the only role of the external electromagnetic fields in this paper will be to provide a driving force on our fluid.

The hydrodynamic variables are number density $n$, temperature $T$ and fluid velocity $v^i$. The fundamental equations governing the dynamics of a non-relativistic fluids are particle conservation, momentum conservation (Navier-Stokes equations) and energy conservation. These three equations read:
\begin{subequations}
\begin{align}
\partial_t n + \partial_i (n v_i)& = 0, \\
\partial_t \left[\varepsilon + \frac{1}{2}nmv_iv_i - \frac{m}{T}\frac{\partial \Pi}{\partial \Omega} \epsilon_{ij}v_i \partial_j T \right] + \partial_i\left[\mathcal{J}_i + \tilde{\mathcal{J}}_i + \mathcal{T}_{ij}v_j \right]  &= E_i J_i,\\
\partial_t (mn v_i) + \partial_j \left(P\delta_{ij} + mnv_iv_j +\tau_{ij} + \mathcal{T}_{ij}\right)& = B\epsilon_{ij}J_j + E_i qn.
\end{align}
\end{subequations}
The terms corresponding to $\partial\Pi/\partial\Omega$, $\mathcal{T}_{ij}$ and $\tilde{\mathcal{J}}_i$ will all correspond to the parity-violating terms in the equations due to the fundamental microscopic constituents of the fluid;  the $B$-field also breaks parity, but is externally imposed in this framework and is not self-consistently determined by the motion of the particles.   Let us explain the physical quantities appearing in the above equations. $\Omega$ denotes the vorticity of the fluid:
\begin{equation}
\Omega = \epsilon_{ij}\partial_iv_j
\end{equation}
 $\varepsilon(n,T)$ is the energy density,
 \begin{equation}
J_i = qn v_i + \frac{q}{T} \frac{\partial \Pi}{\partial \Omega} \epsilon^{ij}\partial_j T.
\end{equation}
denotes the charge current, and
\begin{equation}
\mathcal{T}_{ij} = -\frac{\tilde{\eta}}{2}\left[\epsilon_{ik}\partial_j v_k+\epsilon_{jk}\partial_i v_k+\epsilon_{ik}\partial_k v_j+\epsilon_{jk}\partial_k v_i \right]
\end{equation}
defines the Hall stress tensor.
The curly $\mathcal{J}_i$ and $\tilde{\mathcal{J}}_i$ are the parity preserving and parity violating energy currents respectively:\begin{subequations}\begin{align}
\mathcal{J}_i &= \left(P+\varepsilon + \frac{1}{2}nmv_kv_k\right)v_i + \tau_{ij}v_j - \kappa \partial_i T, \\
\tilde{\mathcal{J}}_i &= q \frac{\partial \Pi}{\partial \Omega} \left(\epsilon_{ij}E_j-Bv_i\right) -\left(\tilde{\kappa}+\frac{v_kv_k}{2T}\frac{\partial \Pi}{\partial \Omega} \right)\epsilon_{ij}\partial_j  T + \frac{m}{T} \frac{\partial \Pi}{\partial \Omega} \epsilon_{ji}v_j \partial_t T
\end{align}\end{subequations}where $\tau_{ij}$ is the usual viscous stress tensor \begin{equation}
\tau_{ij} = - \eta (\partial_i v_j + \partial_j v_i - \partial_k v_k \delta_{ij})  - \zeta \delta_{ij}\partial_k v_k
\end{equation}
For conservation of charge to be true, we require that \begin{equation}
\frac{\partial \Pi}{\partial\Omega} \equiv f(T)
\end{equation} for some function $f$.   Note that in \cite{kaminski}, they split the pressure into two pieces -- in the equations as presented above, this is unnecessary (it is useful, though, for understanding thermodynamic properties).

It is often convenient to use conservation laws to simplify others.   For example, using particle conservation, momentum conservation becomes \begin{equation}
mn (\partial_t v_i + v_j\partial_j v_i) + \partial_i P+ \partial_j (\tau_{ij}+\mathcal{T}_{ij}) = B\epsilon_{ij}J_j + E_i qn
\end{equation}and using particle and momentum conservation, energy conservation becomes \begin{align}
&\partial_t \varepsilon + \partial_i (\varepsilon v_i) + (P\delta_{ij}+\tau_{ij})\partial_i v_j - m\frac{f}{T}\epsilon_{ij}\partial_j T \partial_t v_i + m\frac{f}{T}(\partial_t T) \epsilon_{ij} \partial_j v_i +  q\partial_i \left(f(\epsilon_{ij}E_j - Bv_i)\right) \notag  \\
&= \partial_j \left(\kappa \partial_j T + \left(\tilde{\kappa} + \frac{v_kv_k f}{2T}\right)\epsilon_{ji}\partial_i T\right)+  (E_i-B\epsilon_{ij}v_j) \frac{qf}{T} \epsilon_{ik}\partial_k T \label{eqencons}
\end{align}
As pointed out in the introduction the phenomenology of the solutions of the non-relativistic equations with parity odd terms is not understood. In the next section we will check how the classic solutions of fluid dynamics are affected.

To be absolutely safe, one should only use this theory to compute \emph{first order} (both in gradients and in external electromagnetic fields) corrections to a parity symmetric flow, to ensure that the truncation of the gradient expansion to first order was consistent.   However, we know from practical experience with classical parity-symmetric fluids that there are many realistic experimentally-accessible fluid flows where viscous corrections can dominate over leading order terms:  this is the creep flow or low Reynolds number limit \cite{landau}.   Thus, we will allow ourselves to consider phenomenology when the first-order parity-violating corrections are comparable to the zeroth order dynamics of sound, for example, with the caveat that this approximation may break down.    Another approximation we will almost always make is that the many thermodynamic functions such as $\varepsilon(n,T)$ or $P(n,T)$, and transport coefficients such as $\eta(n,T)$, are approximately constant in the flows of interest.    Finally, we will simply posit boundary conditions on our fluid flows, though it may not be feasible to impose these boundary conditions for some  parity-violating fluids.   These are all complications which we will ignore for a first treatment of the subject.

\subsection{Abbreviations}
Throughout this paper, we will be using some abbreviations for combinations of hydrodynamic transport coefficients, or their derivatives.  Many of these are common in the literature.  Here we collect them for easy reference.

It is often convenient to talk about kinematic viscosities as opposed to normal viscosities, especially in incompressible flows.   These are defined as \begin{subequations}\begin{align}
\nu &= \frac{\eta}{mn}, \\
\nu_\zeta &= \frac{\zeta}{mn}, \\
\tilde{\nu} &= \frac{\tilde{\eta}}{mn}.
\end{align}\end{subequations}We will usually treat these as constants.  In the presence of a magnetic field, there are oscillations at long wavelenghts at the cyclotron frequency \begin{equation}
\omega_{\mathrm{c}} = \frac{qB}{m}.
\end{equation}This can be understood by a simple analogy to the dynamics of a single particle in a magnetic field.

For compressible flows, we define the speed of sound in the usual way: \begin{equation}
c^2 = \frac{1}{m}\frac{\partial P}{\partial n}.
\end{equation}Similarly: \begin{equation}
\alpha \equiv \frac{1}{mn} \frac{\partial P}{\partial T}.
\end{equation}We will define the following two derivatives of $\varepsilon$:  \begin{subequations}\begin{align}
\mathcal{C}_n &\equiv \frac{\partial \varepsilon}{\partial n}, \\
\mathcal{C}_T &\equiv\frac{\partial \varepsilon}{\partial T}.
\end{align}\end{subequations}We will find it useful to define \begin{equation}
\mathcal{E} = \varepsilon + P - qf(T)B
\end{equation}as the coefficient of $\partial_i v_i$ in the energy conservation law.

\section{Effects of Parity Violation on Flux Boundary Conditions} \label{sec2}
As we will see through much of this paper, in many instances parity violation is effectively a ``surface effect" on the hydrodynamic variables $T$ and $v_i$, in incompressible flows.  This does not mean, however, that parity-odd transport coefficients cannot be detected by studying these flows, as one can choose boundary conditions on momentum or energy flux, both of which include parity-violating corrections (see, e.g., \cite{lapa}).   So let us consider the simplest possible toy model of the consequences of parity violation on boundary conditions.  In particular, let us consider the flow of heat in a parity violating fluid, setting $n$ to be constant and $v_i=0$.   A static solution to the equations of motion is simply that the temperature obey's Laplace's equation: \begin{equation}
\partial_j\partial_j T =0.
\end{equation}

Let us consider the following set-up:  imagine that we have a fluid in a cylindrical region, which is periodic in the $x$ direction with  $x\sim x+L_x$, and extends from $0\le y\le L_y$.   We choose the boundary conditions so that we specify $\mathcal{J}_y = -\kappa \partial_y T +\tilde{\kappa} \partial_xT$ on the $y$ boundaries.   Let us see what happens:  define \begin{equation}
T = A_0 + B_0 y + \sum_{n\ne 0} \mathrm{e}^{2\pi\mathrm{i}nx/L_x} \left(A_n \mathrm{e}^{2\pi ny/L_x} + B_n \mathrm{e}^{-2\pi ny/L_x}\right).
\end{equation}To match the boundary conditions, let us suppose that at $y=0$ ($-$) and $y=L_y$ ($+$): \begin{equation}
\mathcal{J}_x(y=0,L_y) = \sum_{n\in\mathbb{Z}} \mathcal{J}^\pm_n  \mathrm{e}^{2\pi\mathrm{i}nx/L_x}.
\end{equation}Since this is a linear equation, we can solve it one Fourier mode at a time.   The zero mode is easiest:  for the boundary conditions to be consistent, we require that \begin{equation}
-\kappa \frac{B_0}{h} = \mathcal{J}^+_0 = \mathcal{J}^-_0,
\end{equation}and the choice of $A_0$ is arbitrary.    For the non-zero modes, we find after an easy calculation of matching boundary conditions that  \begin{equation}
\left(\begin{array}{c} A_n \\ B_n \end{array}\right) = \frac{L_x}{4\pi n \left(\kappa^2+\tilde{\kappa}^2\right)\sinh(2\pi nL_y/L_x)} \left(\begin{array}{cc}  \mathrm{e}^{-2\pi n L_y/L_x} (\mathrm{i}\tilde{\kappa}-\kappa)   &\   \kappa - \mathrm{i}\tilde{\kappa}   \\   -\mathrm{e}^{2\pi n L_y/L_x} (\mathrm{i}\tilde{\kappa}+\kappa)   &\  \kappa + \mathrm{i}\tilde{\kappa}  \end{array}\right) \left(\begin{array}{c} \mathcal{J}^-_n \\ \mathcal{J}^+_n \end{array}\right).
\end{equation}Suppose that we only choose $\overline{\mathcal{J}^\pm_{- n}}=\mathcal{J}^\pm_{ n}$, so that the boundary conditions impose real heat flow.  Comparing $A_n$ with $B_{-n}$ we conclude that $\overline{A_n} = B_{-n}$, so $T$ will be real, as it should.

\begin{figure}[h!]
\centering
\includegraphics[width=3 in]{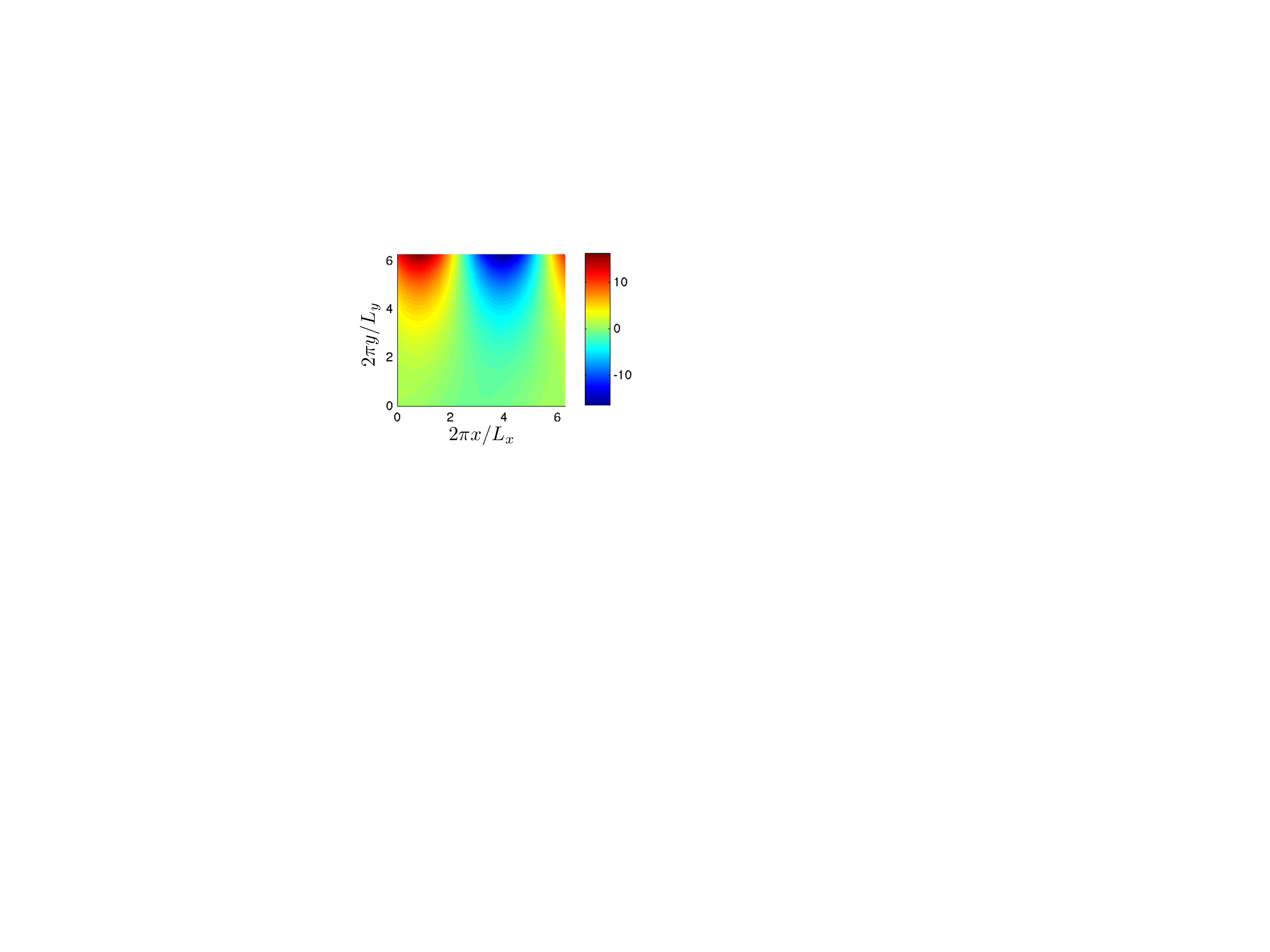}
\caption{An example of Eq. (\ref{tt0}) for $\kappa=\tilde{\kappa}$, $n=1$ and $L_x=L_y$.   The overall normalization of $T$, and the value of $T_0$ are arbitrary.}
\label{tempfig}
\end{figure}

Let us see how this works in a simple example, where we only turn on a pair of Fourier modes $\pm n$ ($n>0$).  We only need to focus on a single matrix equation, as we can simply take the real part as the final answer must be real.  Then we find that \begin{align}
T &= \frac{2L_x}{4\pi n \left(\kappa^2+\tilde{\kappa}^2\right)\sinh(2\pi nL_y/L_x)} \mathrm{Re}\left[(\kappa+\mathrm{i}\tilde{\kappa}) \mathrm{e}^{-2\pi ny/L_x + 2\pi\mathrm{i}nx/L_x} \left(\mathcal{J}^+_n- \mathcal{J}^-_n \mathrm{e}^{2\pi nL_y/L_x} \right) \right. \notag \\
&\left.+(\kappa-\mathrm{i}\tilde{\kappa})\mathrm{e}^{2\pi ny/L_x + 2\pi\mathrm{i}nx/L_x} \left(\mathcal{J}^+_n - \mathcal{J}^-_n \mathrm{e}^{-2\pi nL_y/L_x} \right)\right] + T_0
\end{align}We can handle this one piece at a time, by setting $\mathcal{J}^-_n=0$, and $\mathcal{J}^+_n = J \in \mathbb{R}$.   Then we find that \begin{equation}
T-T_0 = \frac{JL_x}{\pi n \left(\kappa^2+\tilde{\kappa}^2\right)\sinh(2\pi nL_y/L_x)} \left[\kappa \cosh\frac{2\pi ny}{L_x} \cos\frac{2\pi nx}{L_x} + \tilde{\kappa}  \sinh \frac{2\pi ny}{L_x} \sin \frac{2\pi nx}{L_x} \right]    \label{tt0}
\end{equation} In addition to having a temperature gradient near a boundary where there is no heat flow, we see also a relative phase shift between the two terms, proportional to the ratio $\tilde{\kappa}/\kappa$.  An example of Eq. (\ref{tt0}) is plotted in Figure \ref{tempfig}.

Let us now try and solve a slightly more difficult problem.  Let us try to solve Laplace's equation on the domain $L_x/4 \ge |x|$, $0\le y\le L_y$, with the boundary conditions that $T(x=\pm L_x/4) = T_0 \pm \alpha x$, and $\mathcal{J}_y = 0$ at $y=0,L_y$.   To do this, we first write down the trivial solution in the case where the fluid has no parity-odd transport coefficients, and then add on the correction due to parity-odd boundary conditions: \begin{equation}
T = T_0 + \alpha x + T_{\mathrm{PV}}.
\end{equation}$T_{\mathrm{PV}}$ solves Laplace's equation, but with the boundary conditions that $T_{\mathrm{PV}} = 0$ at $x= \pm L_x/4$, and that $\mathcal{J}_y = -\tilde{\kappa}\alpha$ at $y=0,L_y$.   If this problem has a solution which has continuous first derivatives, then we must be able to find it by considering the analogous problem on the periodic domain above, where the heat flow is given by \begin{equation}
\mathcal{J}^\pm(x) = \left\lbrace \begin{array}{ll} -\tilde{\kappa}\alpha &\ |x|<L_x/4 \\ \tilde{\kappa}\alpha &\ \text{otherwise}\end{array}\right.,
\end{equation}which implies that \begin{equation}
\mathcal{J}^\pm_n = \frac{-\tilde{\kappa}\alpha}{2 \pi n}  \left(1-(-1)^n\right).
\end{equation}To see that we can find the solution by doubling the domain, it suffices to note that at the regions where the two domains are joined, continuity is trivially ensured by the Dirichlet boundary conditions, as is continuity of the first derivative, since the two solutions will differ by a minus sign.   By linearity, we thus simply need to sum over the solutions similar to what we found  in Eq. (\ref{tt0}): \begin{align}
T_{\mathrm{PV}} &= \sum_{n\text{ odd}} \frac{-\tilde{\kappa}\alpha L_x}{\pi^2n^2(\kappa^2+\tilde{\kappa}^2)\sinh(2\pi nL_y/L_x)} \times \notag \\
&\left[\kappa  \left(\cosh\frac{2\pi ny}{L_x} - \cosh\frac{2\pi n(L_y-y)}{L_x}\right) \cos\frac{2\pi nx}{L_x} + \tilde{\kappa} \left( \sinh \frac{2\pi ny}{L_x} - \sinh\frac{2\pi n(L_y-y)}{L_x}\right) \sin \frac{2\pi nx}{L_x}\right]
\end{align}But now we find a problem.   For this to be an acceptable solution, we require that $T_{\mathrm{PV}}(x=|L_x|/4)= 0$.   However, this is \emph{not} satisfied.    We conclude that there is no solution expressible as a sum of sines and cosines, which is rather interesting.  We suspect that no time-independent  solution exists at all.   If so, this is quite interesting -- these are physically reasonable boundary conditions (using a combination of thermal baths and insulating walls).

\section{Incompressible Flows}\label{sec3}

Let us begin by studying incompressible flows, where $n$ is a constant.  In this case, particle conservation simplifies to the statement that the velocity is divergenceless: \begin{equation}
\partial_i v_i = 0.  \label{incomp}
\end{equation} In 2+1 dimensions, this leads to the fact that the velocity is the ``curl" of a scalar called the stream function $\psi$: \begin{equation}
v_i = \epsilon_{ij}\partial_j \psi.
\end{equation}Analogous to the introduction of the electrostatic potential, it is often simpler to solve for $\psi$ than to solve for $v_i$, but we should keep in mind that $\psi$ is order $-1$ in derivatives -- it is physically appropriate to have large gradients.   The vorticity is \begin{equation}
\Omega = -\partial_j \partial_j \psi.
\end{equation}
Momentum conservation becomes \begin{equation}
mn (\partial_t v_i + v_j\partial_j v_i)   +   \partial_iP - \eta \partial_j \partial_j v_i - \frac{\tilde{\eta}}{2}\left(\epsilon_{ik}\partial_j\partial_j v_k + \epsilon_{jk}\partial_i\partial_j v_k\right) = qn \left(E_i + B\epsilon_{ij}v_j\right) \label{eq31}
\end{equation}
Note that the $\tilde{\eta}$ term is equivalent to $-\tilde{\eta}\partial_i \Omega$.  This equation (with all terms with $P$, $E_i$ and $B$ combined into a forcing function) was derived in \cite{cai} via the fluid-gravity correspondence.     In fact, this equation is (almost) \emph{equivalent to the parity-symmetric Navier-Stokes equations} in the velocity sector \cite{avron}.  It is easy to see this by acting on both sides of this equation by $\epsilon_{ij}\partial_j$, where we recover \begin{equation}
\partial_t \Omega + v_j \partial_j \Omega + \epsilon_{ik}(\partial_k v_j) \partial_j v_i =\partial_t \Omega + v_j \partial_j \Omega   = \frac{\eta}{mn} \partial_k \partial_k \Omega \end{equation}After some algebra, one can use incompressibility to remove the third term on the left hand side.  We have assumed that the electromagnetic fields are stationary, so that $\epsilon_{ij}\partial_i E_j = 0$.  Remember that the vorticity captures nearly all dynamics of $\psi$ in 2+1 dimensions, up to a harmonic function -- but this means that the only dynamics to $v_i$ that $\Omega$ does not capture is a global $u_i(t)$ contribution, with no spatial dependence.

From the above analysis it is easy to see that if we add the parity breaking terms to Navier-Stokes equation the velocity profile of incompressible flows remains unchanged, so long as the boundary conditions on velocity are unchanged.  However, the fluid can have a modified pressure profile. As an example consider Poiseuille flow \cite{landau}: pressure-driven flow through a channel of length $L$, between two rigid boundaries located at $y=\pm a$. The velocity has only one component $v_x (y)$. We assume that the fluid is driven by a steady preassure drop $\delta P$
\begin{subequations}
\begin{align}
P&=P(x,y),\\
P(0,y)&=P^\ast+\delta P +P(y),\\
P(L,y)&=P^\ast.
\end{align}
\end{subequations}
Moreover, we assume the vanishing velocity at the boundaries of the channel
\begin{equation}
v_x(a)=v_x(-a)=0
\end{equation}
For the above boundary condition the velocity profile has the following form
\begin{equation}
v_x(y)=\frac{\delta P}{2 \eta L }(a^2-y^2)  \label{eq35}
\end{equation}
Note especially that the unperturbed velocity field is independent of $x$, and that the unperturbed
pressure gradient is constant in the $x$ direction. This velocity is the same as for parity preserving flows. However, in the case of parity breaking flows the fluid develops gradient in the $y$ direction
\begin{equation}
P'(y)\neq 0
\end{equation}
which is a novel effect coming from symmetry breaking. The explicit pressure profile can be found by plugging Eq. (\ref{eq35}) into Eq. (\ref{eq31}):
\begin{equation}
P(x,y)= P_0 + \delta P \left(1-\frac{x}{L}\right) -\frac{2 \delta P \tilde{\eta}  y}{\eta  L}
\end{equation}
From that expertise we conclude that in general, parity-violation creates new non-trivial pressure profiles in the fluid.     It is worth noting that this solution does not satisfy the physically reasonable boundary conditions that the pressure $P$ is a constant along the boundaries $x=0$ and $x=L$.   It may be the case that there is another solution to the (nonlinear) Navier-Stokes equations that is consistent with all boundary conditions, including that the pressure at the ends of the channels is $y$-independent, but we have not found such a solution.


\subsection{Flow Down an Inclined Plane}

\begin{figure}
\centering
\includegraphics[width=0.55\textwidth]{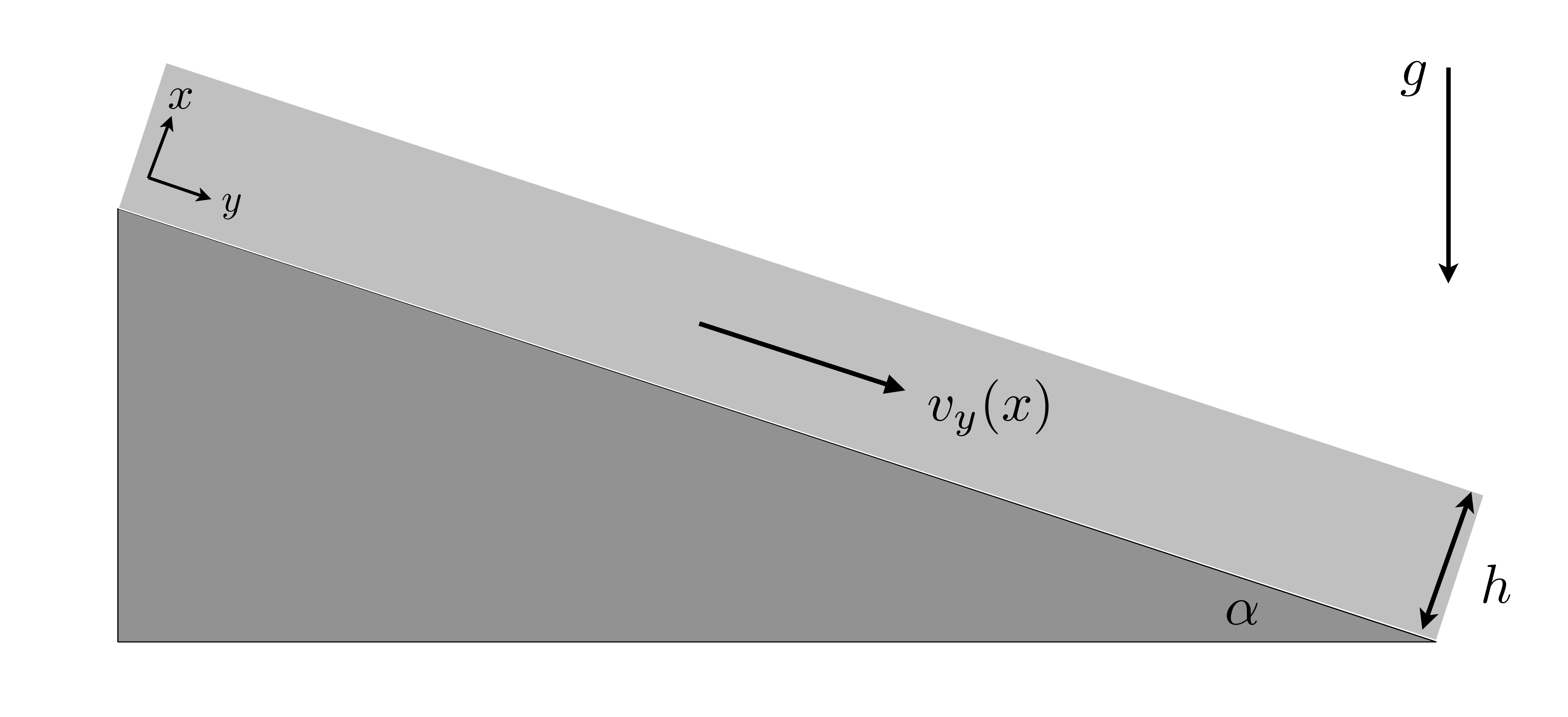}
\caption{The set-up of flow on an inclinded plane.}
\label{inclined}
\end{figure}

A distinguishing feature of fluids with non-zero Hall viscosity is the stress that is exerted in the direction perpendicular to the usual viscous stresses. Therefore it is natural to expect that interesting physical consequences may occur if we create an imbalance between these stresses in the fluid. One example of a system with such imbalance is a two-dimensional fluid film flowing down an ``inclined plane" at an angle $\alpha$, due to ``gravitational force"  \cite{Bruus2008}:  see Figure \ref{inclined}.  We stress here that this flow truly exists in an effective two-dimensional space -- translation invariance in the third direction would alter the parity-violating gradient expansion.  This gravitational field at an angle can easily be achieved in a charged parity-violating fluid by turning on appropriate electric fields.   We label the thickness of the film $h$.
The flow is driven by the gravitational acceleration $\vec{g}$, with components $g_x = g\cos\alpha$, $g_y=g\sin\alpha$. The $x$ component of the gravitational force will be balanced by the force due to the Hall viscosity. The component parallel to the plane accelerates the film down
along the inclined plane until the velocity of the film is so large that the associated viscous
friction forces in the film compensate $v_y$. When this happens the motion of the film has
reached a steady state, as shown in Figure \ref{inclined}. The translation invariance of the setup along the $y$ directions dictates that the
velocity field can only depend on $x$. We demand two boundary conditions, which are given by requiring no-slip of $v$ at the plane $x = 0$ and no stress on the free surface
\begin{equation}
v _y (0)=0
\end{equation}
\begin{equation}
\partial _x v _y (h) =0
\end{equation}
The $y$ component of the velocity profile remains unmodified
\begin{equation}
v_y (x) =\frac{g mn_0  \sin (\alpha ) \left(2 h x-x^2\right)}{2 \eta }
\end{equation}
However, we obtain a non-trivial correction to the pressure distribution in the fluid
\begin{equation}
P(x)=P^\ast +mg n_0  (h-x) \left(\cos (\alpha )+ \frac{ \tilde{\eta }}{\eta} \sin (\alpha )\right)
\end{equation}
where $P^\ast$ is the integration constant, which we take to be defined at $x=h$, as the external pressure at the free surface.   As one can check explicitly, the combination of $P+\tau_{xx} + \mathcal{T}_{xx}$ is $\tilde{\eta}$-independent.   This corresponds to the requirement that Newton's Laws be satisfied at all points in the fluid.   In particular, this means that the stresses at the interface $x=0$ are not altered by Hall viscosity.   Curiously, at a special value of $\tilde{\eta}/\eta = -\cot\alpha$, the pressure gradients vanish.

In the transverse direction, the total net stress tensor component $\sigma_{yy} = P+\tau_{yy}+\mathcal{T}_{yy}$ is not independent of $\tilde{\eta}$: \begin{equation}
P+\tau_{yy}+\mathcal{T}_{yy} = P^\ast + mg n_0  (h-x) \left(\cos (\alpha )+2 \frac{ \tilde{\eta }}{\eta} \sin (\alpha )\right).
\end{equation}This may have interesting implications: for example, suppose that the flow is not quite translation invariant in $y$, and there is some edge of the fluid (perhaps held in place by surface tension).   Due to the change in the internal stresses acting on the fluid near this edge, we expect the shape of this boundary to be altered.  In particular, curious effects may happen if we take $\cos\alpha + 2\sin \alpha \tilde{\eta}/\eta <0$, in which case the ``hydrostatic" pressure actually pulls the fluid inward.   Alternatively, if we consider a film with a slowly varying height $h(y)$, when $\tilde{\eta}$ is small then imbalances in $h(y)$ will tend to be corrected, as hydrostatic pressure will tend to push out regions of high $h$ into regions of small $h$.  Curiously, Hall viscosity can evidently \emph{reverse} this effect, suggesting that this flow becomes unstable.

\subsection{Couette Flow with a Temperature Gradient}
Next, consider a Couette flow with a temperature gradient.  We look for solutions which are independent of $t$ and $x$, in between plates at $y=0,h$ with boundary conditions that $T=T_0$ and $v_x=0$ at $y=0$, and $T=T_1$ and $v_x = v_0$ at $y=h$, and $v_y=0$ at both.   The equations of motion for an incompressible flow are (straightforwardly) $v_y=0$ (from particle conservation), and \begin{subequations}\begin{align}
-\nu \partial_y^2 v_x &= 0, \\
\partial_y P + \tilde{\eta} \partial_y^2 v_x  &= -qBn_0 v_x - BG\partial_y T, \label{eq40b} \\
\frac{\eta}{2}(\partial_y v_x)^2  &= \kappa \partial_y^2 T
\end{align}\end{subequations}Note that we have turned on a magnetic field $B$ in order to observe parity-violating effects.  We have defined $G\equiv qf(T)/T$, and assumed that $E_i=0$.    It is easy to see that $v_x$ and $T$ are un-altered by parity-violation:\begin{subequations}\begin{align}
v_x &= \frac{v_0y}{h},\label{eq41a} \\
T &= T_0 + \frac{T_1-T_0}{h}y + \frac{\eta v_0^2}{4\kappa h^2}y(y-h).  \label{eq41b}
\end{align}\end{subequations}However, the pressure will now be non-constant: \begin{equation}
P=P_0  - \frac{qn_0Bv_0}{2h}y^2 - BG\left(\frac{T_1-T_0}{h}y + \frac{\eta v_0^2}{4\kappa h^2}y(y-h)\right).
\end{equation}In particular, there will be now a pressure difference $\Delta P$ between the top and bottom plates: \begin{equation}
\Delta P = -BG\Delta T - \frac{qnB v_0h}{2}.
\end{equation}Note that this flow leads to \emph{different} boundary conditions on the fields and stresses than the parity-symmetric flow.   This is not the case in our examples without thermal flows.\footnote{This can be seen without further calculation -- the fluid stresses $\tau_{ij}$ and $\mathcal{T}_{ij}$ are $B$-independent, but $P$ is.}

Even in the presence of a magnetic field, it is straightforward to see from Eq. (\ref{eq40b}) that the effects of temperature can be re-absorbed into an effective pressure (much like the Hall viscosity).   This is even true if we allow for arbitrary dependence on $t$, $x$ and $y$.   This means that, for example, if we choose Couette flow boundary conditions, then Eq. (\ref{eq41a}) is always valid.   However, we can now consider non-trivial dynamics of temperature: \begin{equation}
\mathcal{C}_T (\partial_t T + v_x \partial_x T) - \frac{\eta v_0^2}{2h^2}  + \frac{mfv_0}{Th}\partial_t T + qBv_x \partial_x f(T) = \kappa \partial_j \partial_j T + \frac{v_x v_0 f(T)}{hT}\partial_x T - Bv_x \frac{qf}{T} \partial_x T \label{eq45}
\end{equation}If we assume that $\partial_xT=0$, the \emph{only} change to this equation is as follows.   Let $T_{\mathrm{stat}}$ be the solution given by Eq. (\ref{eq41b}).   It is easy to see that all of the convective terms will drop out of Eq. (\ref{eq45}).   If we let $T=T_{\mathrm{stat}}+ \delta T$, then: \begin{equation}
\left(\mathcal{C}_T + \frac{mv_0G}{qh}\right)\partial_t \delta T = \kappa \partial_y^2 \delta T.
\end{equation}We see that the effective heat capacity is dependent on the value of $v_0$.  This is because the planar Couette flow leads to a constant vorticity, and parity-violation leads to an explicit $\Omega$-dependent correction to the heat capacity.

\subsection{Waveguide with an Interface}
In this subsection, we will compute the eigenfrequencies of a ``waveguide" for a channel, in which the top half contains a parity-even incompressible fluid, and the bottom half contains a parity-odd incompressible fluid.   Of course, a typical waveguide contains a compressible fluid.  Our main point in this section is to point out that the presence of a boundary where the Hall viscosity jumps can lead to a rather exotic effect on the modes of the waveguide.  For this purpose, it will suffice to ignore the propagation of sound waves, and consider the simpler problem where the fluids are incompressible and viscous.

In a parity-even fluid, if we find a mode of a waveguide with a given $\omega$ and $k$, there must be a mode with the same value of $\omega$, but the opposite wave number $-k$; namely, the dispersion spectrum of the system is parity-even.   For example, with simple sound waves, we have $\omega = \pm ck$.

Now, imagine that we have an incompressible fluid placed in between two rigid plates at $y=\pm h/2$, and that the (kinematic) Hall viscosity is \begin{equation}
\tilde{\nu}(y) = \tilde{\nu}\;\mathrm{sign}(y).
\end{equation}Because the system is translation invariant in $x$ and $t$, we are free to look for solutions where the stream function is of the form $\psi = \psi(y)\mathrm{e}^{\mathrm{i}kx-\mathrm{i}\omega t}$.  Although for $y>0$ and $y<0$, the effects of the Hall viscosity can clearly be ignored, the presence of an interface at $y=0$ means that there will be chiral corrections to the ``dispersion relation" $\omega(k)$ for ``propagating" waves down the channel.    Without Hall viscosity, in fact such waves will simply decay, but we will see that this becomes a true (dissipative) waveguide when we account for Hall viscosity.

For simplicity, let us work in the low Reynolds number limit, where we can neglect the nonlinear convective term in the momentum conservation equations.  After a straightforward calculation, we find that the momentum conservation equations combine into a single equation describing $\psi(y)$: \begin{equation}
\mathrm{i}\omega \left(k^2-\partial_y^2\right)\psi = \nu \left(k^2-\partial_y^2\right)^2\psi - 2\mathrm{i}\tilde{\nu}k^3 \delta(y) \psi.  \label{wgeq}
\end{equation}Away from $y=0$, the solution to this equation is a linear combination of exponentials of the form $\mathrm{e}^{\pm ky}$ and $\mathrm{e}^{\pm qy}$, with \begin{equation}
q = \sqrt{k^2 - \frac{\mathrm{i}\omega}{\nu}}.
\end{equation}

In principle, one could match boundary conditions at $y=0$ and solve an 8-dimensional linear algebra problem to find $\omega$;  in practice, this cannot be done analytically, except perturbatively at small $\tilde{\nu}$, as we explain in Appendix \ref{appa}.   Nonetheless, the differential equations above, with the boundary conditions $ \psi = \partial_y \psi = 0$ (corresponding to $v_i=0$) at the boundaries $y=0,h$, form a generalized eigenvalue problem which is well suited for accurate numerical solutions using pseudospectral methods based on expanding $\delta \psi$ in terms of $N$ Chebyshev polynomials \cite{trefethen, huang}.    We implemented the surface boundary condition numerically by implementing the $\delta$ function as constant functions over an O(1) number of grid points, of height $N$.   In all of the numerical results displayed in this paper, we used $N=53$.    The result was robust to the precise resolution of $\delta(y)$ and to the value of $N\gg 1$.

\begin{figure}[h!]
\centering
\includegraphics{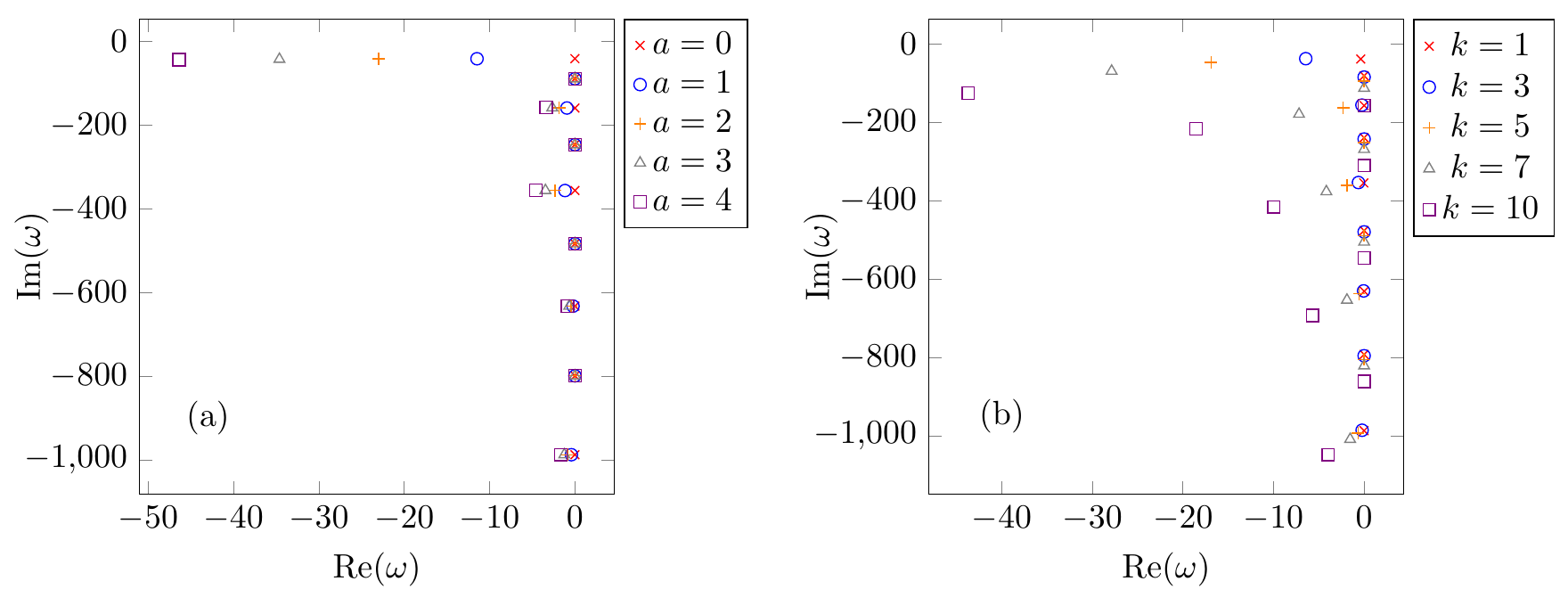}
\caption{The 9 most unstable modes of the parity-violating waveguide. (a): $k=4$ with varying $a$.   (b): $a=1$ with varying $k$.}
\label{waveeigfig}
\end{figure}

It is helpful to non-dimensionalize $k$ by writing $k=\hat{k}h$, and $\omega$ by $\omega = \hat{\omega}\nu h^{-2}$.    We also define the dimensionless parameter \begin{equation}
a \equiv \frac{\tilde{\nu}}{\nu}.
\end{equation}(From now on we drop the hats in this section.)   Evidently, there are universal dispersion relations, characterized by a single dimensionelss number $a$ corresponding to the strength of parity violation at the interface.    In Figure \ref{waveeigfig}, we have plotted the real and imaginary parts of $\omega(k)$ for select values of $a$ and $k$, for some of the most unstable modes.   Half of the modes are essentially unaltered by parity-violation -- these are odd modes which vanish at the interface, as we explain in the appendix.    In Figure \ref{waveRIfig}, we show both the real and imaginary part of $\omega(k)$ as a function of $k$, for the most unstable mode.   For small $k$, essentially the only change to the dispersion relation is that $\mathrm{Re}(\omega) \sim -ak^3$, as explained in the appendix.   These waves are not localized on the interface -- they simply correspond to the lowest normal mode.   For larger $k$, evidently this behavior breaks down.

\begin{figure}[h!]
\centering
\includegraphics{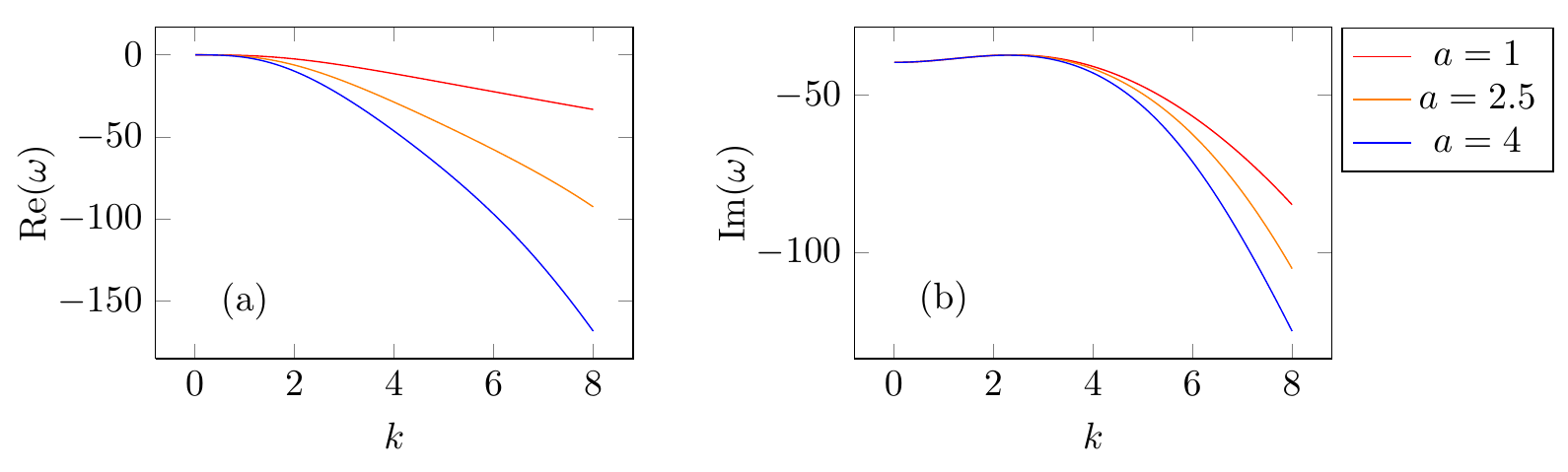}
\caption{The (a) real and (b) imaginary parts of the most unstable mode $\omega(k)$, for various values of $a$.}
\label{waveRIfig}
\end{figure}

The main observation of this section is the observation of parity-violation in the dispersion spectrum of this waveguide.   For example, the modes with $\mathrm{Re}(\omega) \sim -ak^3$ pick out a preferred direction -- waves always propagate to the left (if $a>0$) along the interface.   We expect that this effect persists if we allow the fluids to be compressible.

In fact, waveguides such as this may be an interesting way  of detecting parity-violation in a laboratory.   If the parity-violating contributions are ``small", it may be possible to observe slow beating effects between the frequencies of the left-moving and right-moving modes in the waveguides, whose $\mathrm{Re}(\omega)$ are slightly altered by the presence of the ``parity-violating interface."

\section{Sound Waves}\label{sec4}
Let us begin our study of compressible flows by studying the propagation of sound waves.   Starting with a fluid at rest, let us imagine perturbing it by \begin{subequations}\begin{align}
n &= n_0 + \delta n, \\
v_i &= \delta v_i, \\
T &= T_0 + \delta T,
\end{align}\end{subequations}and study perturbations of the form $\mathrm{e}^{\mathrm{i}(kx-\omega t)}$ in an unforced system. The conservation laws of particles, momentum, and energy become, in an external magnetic field $B$:  \begin{subequations}\begin{align}
\mathrm{i}\omega \delta n &= \mathrm{i}k n_0 \delta v_x, \\
\mathrm{i}\omega \delta v_x &= \mathrm{i}k\frac{c^2}{n_0}\delta n +  (\nu+\nu_\zeta) k^2 \delta v_x + \left(\tilde{\nu}k^2-\omega_{\mathrm{c}}\right)\delta v_y + \mathrm{i}k(\alpha+BG)\delta T, \\
\mathrm{i}\omega \delta v_y &= \nu k^2\delta v_y - \left(\tilde{\nu} k^2-\omega_{\mathrm{c}}\right)\delta v_x , \\
-\mathrm{i}\omega \mathcal{C}\delta T -\mathrm{i}\omega \mathcal{C}_n\delta n + \mathrm{i}k\mathcal{E}\delta v_x &= -\kappa k^2\delta T.
\end{align}\end{subequations}where again, $G=qf(T)/T$.  It is easy to combine these to find the dispersion equation: \begin{equation}
\omega = \frac{c^2k^2}{\omega} - \mathrm{i}(\nu+\nu_\zeta) k^2 + \frac{\left(\tilde{\nu}k^2-\omega_{\mathrm{c}}\right)^2}{\omega + \mathrm{i}\nu k^2}  +\frac{\lambda k^2}{\mathcal{C}_T\omega  +\mathrm{i} \kappa k^2}
\end{equation}where \begin{equation}
\lambda \equiv (\mathcal{E}-\mathcal{C}_nn_0)(BG + \alpha).
\end{equation}    The first thing we must do is determine the long-time, long-wavelength modes.  A careful analysis of this equation leads to the following four modes, if the background magnetic field $B=0$: \begin{equation}
\omega = \pm \sqrt{c^2 + \frac{\lambda}{\mathcal{C}_T}} k, \;\; -\mathrm{i} \nu k^2, \;\; -\mathrm{i}\frac{\kappa}{\mathcal{C}_T - \lambda c^{-2}}k^2.   \label{eq57}
\end{equation}There are two (small) parity-violating corrections:  one to the speed of sound, and one to the thermal diffusion constant.   If, however, $B \ne 0$, then we find a very different limit: \begin{equation}
\omega = \pm \omega_{\mathrm{c}},\;\; -\frac{\mathrm{i}\nu c^2}{\omega_{\mathrm{c}}^2}k^4,\;\;-\mathrm{i}\frac{\kappa}{\mathcal{C}_T}k^2. \label{eq58}
\end{equation} In particular, we note the presence of a very long lived diffusive mode with $\omega \sim k^4$.     We emphasize that there is an important order of limits at play -- if $\omega / \omega_{\mathrm{c}} \rightarrow \infty$, before $\omega \rightarrow 0$, we obtain Eq. (\ref{eq57});  if $\omega/\omega_{\mathrm{c}}\rightarrow 0$ first, we obtain Eq. (\ref{eq58}).   We can also study the sound modes  as $\omega,\omega_{\mathrm{c}},k \rightarrow 0$ at the same rate.   For example, the sound modes become $\omega = \pm \sqrt{\omega_{\mathrm{c}}^2 + (c^2+\lambda/\mathcal{C}_T)k^2}$.

It is straightforward to see that, in general, there is no instability of a fluid at rest.   To do this, we look for dispersive modes with real $\omega$ -- this is because the dispersion relation is a smooth function of $\tilde{\eta}$, we know that all modes are in the lower half plane when $\tilde{\eta}=0$, and any unstable mode must therefore, at some point, cross the real $\omega$ axis.   If $\omega$ is real, then we must satisfy the equation \begin{equation}
(\nu+\nu_\zeta)k^2 + \frac{\nu k^2 \left(\tilde{\nu}k^2-\omega_{\mathrm{c}}\right)^2}{\omega^2 + (\nu k^2)^2} + \frac{\lambda \kappa k^4}{(\mathcal{C}_T\omega)^2 + (\kappa k^2)^2}=0
\end{equation} which is clearly impossible unless $\nu=\nu_\zeta = \kappa =0$, or $k=0$.

If we take $\eta=\zeta=\kappa=0$ (this describes a non-dissipative fluid), then we get the dispersion relation \begin{equation}
\omega = \pm\sqrt{\left(c^2 + \frac{\lambda }{\mathcal{C}_T}\right)k^2 + \left(\tilde{\nu} k^2 - \omega_{\mathrm{c}}\right)^2 }.
\end{equation}This is reminiscent of the propagation of helicon waves \cite{maxfield} at large $k$, where $\omega \sim k^2$.  These are non-decaying, instantly propagating waves, analogous to the non-relativistic free particle in quantum mechanics.    This is not surprising, as the parity-violating  terms are non-dissipative.    We have, so far, only found two modes.    The remaining two modes are non-dynamical:  $\omega=0$.

\subsection{Driving a Plate}
As an interesting example of the consequences of these sound modes, let us consider the set-up where we have a plate at $y=0$ which is free to slide in the $x$-direction (boundary condition $\sigma_{xy}=0$), and a rigid plate at $y=h$ which oscillates up and down with a very small velocity $v_{\mathrm{drive}} \cos(\omega t)$.    For simplicity, we assume there is translation invariance in the $x$-direction, and that $B=0$.   In a parity-symmetric fluid, such a flow would induce pressure oscillations in between the two plates, but because there is a symmetry under $x \rightarrow -x$ that is preserved by the driving force, there will be no motion in the $x$ direction of the plate at $y=0$.    However, we expect that a parity-odd fluid will have motion of the $y=0$ plate, because the spontaneous drive of the $y=h$ plate in the $y$ direction breaks the symmetry $x\rightarrow -x$.   We will now show that this is indeed the case.  Note that the mechanism behind the motion of the plate in the $x$-direction are the stresses induced on the bottom plate by the Hall viscosity.

\begin{figure}[h!]
\centering
\includegraphics{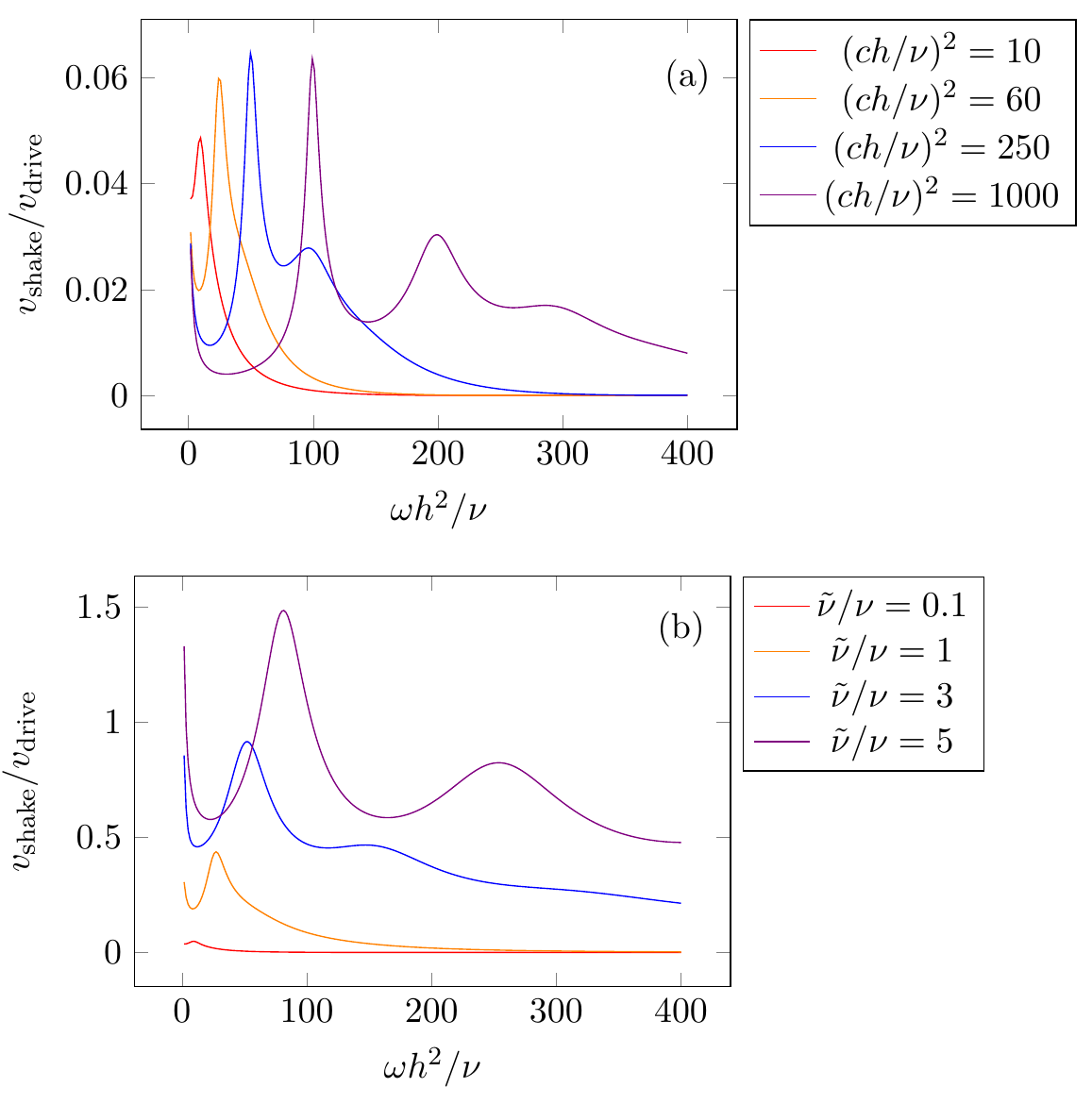}
\caption{(a) $v_{\mathrm{shake}}/v_{\mathrm{drive}}$ with fixed $\tilde{\nu}/\nu=0.1$.   (b)  $v_{\mathrm{shake}}/v_{\mathrm{drive}}$ with fixed $(ch/\nu)^2 = 60$.}
\label{shakefig}
\end{figure}

A very similar calculation to the above calculation of sound modes reveals that there are 4 possible modes of oscillation, which are of the form $\mathrm{e}^{\mathrm{i}ky-\mathrm{i}\omega t}$, where $k=\pm k_{\pm}$ with \begin{equation}
k^2_\pm \equiv \frac{-2\nu \omega^2 - \mathrm{i}\omega c^2 \pm  \mathrm{i}\sqrt{\omega^2 c^4 + 4\tilde{\nu}^2 \omega^4}}{2\left(\left(\nu^2+\tilde{\nu}^2\right)\mathrm{i}\omega - \nu c^2\right)},
\end{equation}where we have at momentum $k$:\begin{equation}
\delta v_y = \left(\frac{\mathrm{i}\omega}{\tilde{\nu} k_\pm^2} - \frac{\nu}{\tilde{\nu}}\right)\delta v_x \equiv \gamma_{\pm} \delta v_x.
\end{equation}Writing the solution as (the real part of) \begin{equation}
\delta v_x = \left(A\mathrm{e}^{\mathrm{i}k_+y}+B\mathrm{e}^{-\mathrm{i}k_+y}+ C\mathrm{e}^{\mathrm{i}k_-y}+D\mathrm{e}^{-\mathrm{i}k_-y} \right)\mathrm{e}^{-\mathrm{i}\omega t},
\end{equation}after some algebra we find we simply have to solve the linear algebra problem: \begin{equation}
\left(\begin{array}{cccc} k^{-1}_+ &\ -k^{-1}_+ &\ k^{-1}_- &\ -k^{-1}_- \\ \gamma_+ &\ \gamma_+ &\ \gamma_- &\ \gamma_- \\ \mathrm{e}^{\mathrm{i}k_+h} &\ \mathrm{e}^{-\mathrm{i}k_+h} &\ \mathrm{e}^{\mathrm{i}k_-h} &\ \mathrm{e}^{-\mathrm{i}k_-h} \\ \gamma_+\mathrm{e}^{\mathrm{i}k_+h} &\ \gamma_+\mathrm{e}^{-\mathrm{i}k_+h} &\ \gamma_-\mathrm{e}^{\mathrm{i}k_-h} &\ \gamma_-\mathrm{e}^{-\mathrm{i}k_-h} \end{array}\right) \left(\begin{array}{c} A \\ B \\ C \\ D \end{array}\right)=\left(\begin{array}{c} 0 \\ 0 \\ 0 \\ v_{\mathrm{drive}} \end{array}\right)
\end{equation}The first line corresponds to the boundary condition $\sigma_{xy}=0$, which simplifies a lot.   The magnitude of shaking of the bottom plate is given by  \begin{equation}
|A+B+C+D| \equiv v_{\mathrm{shake}} = \frac{v_{\mathrm{drive}}}{2}\left| \frac{k_-\left(\mathrm{e}^{\mathrm{i}k_-h}-\mathrm{e}^{-\mathrm{i}k_+h}\right)-k_+\left(\mathrm{e}^{\mathrm{i}k_+h}-\mathrm{e}^{-\mathrm{i}k_-h}\right)}{\gamma_+ k_+ \sin(k_+h)\cos(k_-h) - \gamma_-k_-\sin(k_-h)\cos(k_+h)}\right|.  \label{vshake}
\end{equation}Note that $\gamma_\pm \sim \tilde{\nu}^{-1}$, so in the limit of small $\tilde{\nu}$, the shaking amplitude becomes proportional to $\tilde{\nu}$ as expected.     We have plotted $v_{\mathrm{shake}}/v_{\mathrm{drive}}$ in Figure \ref{shakefig}.   We can see the emergence of resonances which are suppressed either as the speed of sound, or the Hall viscosity, become small, although we do not have asymptotic control over Eq. (\ref{vshake}) to explain these features.

\section{Dissipationless Flows}\label{sec5}
In this section, we will make an unjustified assumption -- namely, that the dissipative terms $\eta=\zeta=\kappa=0$, but that $\tilde{\eta}\ne 0$.   In doing so, we will find a variety of new solutions to the hydrodynamic equations that are unique to parity-odd fluids.   Many aspects of these flows may be relevant even with dissipative corrections.   But our main motivation here is simply to explore fluid dynamical flows where parity violation causes dramatic effects that are more subtle than the simple emergence of ``Hall stress" at the boundary of the fluid.   We will see new types of solitons arising, as well as the stabilization of flows that are unstable in parity symmetric fluids.   This section is meant to simply give a flavor for the types of uniquely parity-violating fluid flows that we have found, though they may not be the most easily observable examples in experiments.

\subsection{Solitons}
In this subsection, we will look for soliton-like solutions to the fluid equations of motion.   Solitons are a famous class of solutions to nonlinear differential equations that correspond to traveling waves with a time-independent profile (up to a simple translation): i.e., \begin{equation}
n = n(x-vt), \;\;\; u_i = u_i(x-vt).
\end{equation}  This is non-trivial as the wave equations obeyed by solitons are nonlinear, and so in general the solitonic profile is unique.   Solitons arise in many nonlinear systems, and they have been observed in many experiments in hydrodynamics and beyond \cite{remoissenet}.   Here we show that parity-violating solitons can arise when dissipation is weak.

Motivated by the existence of the frozen sound modes when $\eta=\zeta=\kappa=0$, but $\tilde{\eta} \ne 0$, we are tempted to look for an exact solution to the hydrodynamic equations which depends only on one spatial direction (say $y$).   Let us begin by fixing $v_y=0$.      It is easy to see that the only equation which is not automatically satisfied is the $x$-component of momentum conservation, which leads to the constraint \begin{equation}
P(n,T) - P_0 = \tilde{\nu} n_0 \partial_y v_x
\end{equation}where $P_0$, $n_0$ and $T_0$ are arbitrary constants.     This is a solution to the fully nonlinear equations with $\eta=\zeta=0$, with all hydrodynamic constants assumed independent of $n$.      This allows us to construct simple analogues of parity-violating compressible Poiseuille and planar Couette flows, regardless of the form of $P(n,T)$.    We fix the residual freedom in choosing $n$ and $T$ by the constraint that the energy density is a constant.

We can find an exotic generalization of this flow if we allow for particle flux in the $y$ direction, but still require all fields are functions of $y$.  For now, we assume there is no background magnetic field -- we will add one in a later subsection.   Let us begin by assuming that temperature fluctuations are negligible, and as before that we can approximate $\Delta P = mc^2 \Delta n$, over a large range of $n$.  In this case, particle conservation leads to \begin{equation}
nv_y = \text{constant} \equiv J .
\end{equation}Without loss of generality, assume $J>0$.  The $x,y$-component of momentum conservation read respectively \begin{subequations}\label{eq82}\begin{align}
v_y \partial_y v_x &= \tilde{\nu} \partial_y^2 v_y, \\
v_y \partial_y v_y + \frac{c^2}{n} \partial_y n &= -\tilde{\nu} \partial_y^2 v_x.  \label{eq62b}
\end{align}\end{subequations}These equations can be combined: \begin{equation}
v_y \partial_y v_y -c^2\frac{\partial_yv_y}{v_y} = -\tilde{\nu}^2 \partial_y \frac{\partial_y^2 v_y}{v_y} \label{vy2}
\end{equation}which can be non-dimensionalized by rescaling: \begin{subequations}\begin{align}
v_y &\equiv \hat{v}_y  c,\\
y &\equiv \hat{y}l = \hat{y} \frac{\tilde{\nu}}{c}
\end{align}\end{subequations} -- there is a universal family of solutions when $J\ne 0$, with $v_y \sim c f(y/l)$.   Note that this family is only non-trivial when $\tilde{\nu} \ne 0$.

\begin{figure}[h!]
\centering
\includegraphics{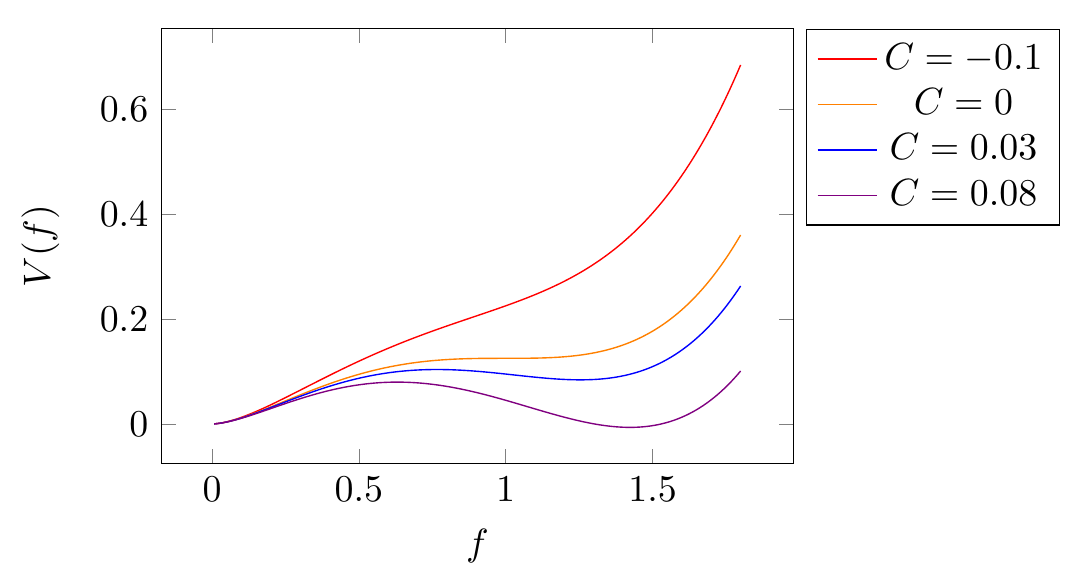}
\caption{A plot of the function $V(f)$ for various choices of $C$.}
\label{vfig}
\end{figure}

$v_y=\text{constant}$ is still a solution.   To find the non-trivial solutions, we notice that Eq. (\ref{vy2}) is a total-derivative:  the new solutions with non-constant $f$ correspond to non-trivial solutions to the equation \begin{equation}
f^{\prime\prime} = f(\log f +C_0) - \frac{f^3}{2}
\end{equation}where $C_0$ is some integration constant.   Solutions to this equation correspond to flows which do not have an analogue in parity-symmetric hydrodynamics.   This equation can be qualitatively solved in analogy to Newton's Second Law -- thinking of the rescaled $y$ as time, and the $f$ as a position, we can write this as \begin{equation}
f^{\prime \prime} = -\frac{\partial V}{\partial f}   \label{eqnewton}
\end{equation} where
\begin{equation}
V(f) = \frac{f^4}{8} - f^2 \left(\frac{\log f}{2}+C\right).
\end{equation}
We have chosen a new integration constant $C$, without loss of generality.  This analogy makes it very easy to understand the dynamics without an analytic solution.   For generic initial conditions, we see that our problem is ill-behaved -- we must stay trapped in the region $f>0$, but this potential is smallest when $f<0$.   To find a well-behaved solution, we thus have to choose $C$ to be large enough that there is a trapping region of the potential at positive $f$.  It is easy to check analytically that in fact, we must choose $C$ to be positive:  see Figure \ref{vfig}.   Finally, we note that using the Newtonian analogy, we know how to integrate Eq. (\ref{eqnewton}) to a first order differential equation, by invoking conservation of ``energy".

Suppose that we have such a trapping region in $V$.  It is easy to see that (nonlinear) oscillatory solutions are possible in space which are periodic with some (tunable) period $\sim l$.   More interestingly, there is a ``kink" solution where at $y=\pm \infty$:  $v_y$ will asymptote to a constant, but around $y=0$, the velocity will flow to some larger value.   We simply think of a particle in this fictitious potential which starts just at this local maximum, moves to a larger value of $f$ near $y=0$, then rolls back, approaching the local maximum exponentially quickly in $y$ at large $|y|$.    We note that this kink solution is unstable -- again using the Newtonian analogy, a kick near $y=\pm\infty$ will generically send the flow towards $f=0$, which leads to an unphysical solution (or more realistically, a breakdown of our assumptions).

What happens if we turn on a small (kinematic) viscosity $\nu$?   Now, we have to modify the velocity equations by adding $\partial_t v_i - \nu\partial_j\partial_j v_i$ to the left hand side.   If we work in a perturbative limit where $f(y)$ is oscillating around the local minimum of $V$, then we can approximate that $f(y) \approx f_0 + f_1 \cos(\alpha y/l)$ where $\alpha$ is an O(1) constant.    We now guess that the time-dependent solution is simply to replace $f(y)$ with $f_0+f_1(t)\cos(\alpha y/l)$.   This will certainly be a solution to Eq. (\ref{eq82});  for the time-dependent piece to solve the viscous piece, we evidently simply need to set $f_1(t) = f_{10} \exp[-t/t_0]$ where \begin{equation}
t_0 \equiv \frac{1}{\alpha^2 \nu l^2} = \frac{c^2}{\alpha^2 \tilde{\nu}^2\nu}
\end{equation}Such a kink solution is observable on time scales smaller than $t_0$.

Next, let us add the fluctuations of temperature to our solution.   This will alter Eq. (\ref{eq62b}) to \begin{equation}
v_y \partial_y v_y + \frac{c^2}{n} \partial_y n + \alpha \partial_y T = -\tilde{\nu} \partial_y^2 v_x.
\end{equation}Energy conservation becomes \begin{equation}
0=v_y\partial_y \varepsilon + \mathcal{E} \partial_y v_y = \mathcal{C}_n v_y \partial_y n + \mathcal{C}_T v_y \partial_y T + \mathcal{E} \partial_y v_y
\end{equation}For simplicity, let us assume that $\mathcal{C}_n$, $\mathcal{C}_T$  and $\mathcal{E}$ are constants, so that we can explicitly find a new effective potential by hand.   The differential equation for $v_y$ becomes\begin{equation}
\partial_y \left(\frac{v_y^2}{2} - \left(c^2 + \frac{\alpha\mathcal{E}}{\mathcal{C}_T}\right)\log \frac{v_y}{c} - \frac{\alpha \mathcal{C}_nJ}{\mathcal{C}_Tv_y}\right) = -\tilde{\nu}^2 \partial_y \frac{\partial_y^2v_y}{v_y}.
\end{equation}Defining $\sqrt{c^2+\alpha\mathcal{E}/\mathcal{C}_T} \equiv c_{\mathrm{eff}}$, the new dimensionless parameter \begin{equation}
A \equiv \frac{\alpha \mathcal{C}_n J}{\mathcal{C}_Tc_{\mathrm{eff}}},
\end{equation}non-dimensionalizing velocities by rescaling by $c_{\mathrm{eff}}$, and normalizing lengths by $\tilde{\nu}/c_{\mathrm{eff}}$, we find the effective potential for $v_y = f$: \begin{equation}
V(f) = \frac{f^4}{8} - f^2 \left(\frac{\log f}{2}+C\right) - Af.
\end{equation}

One can certainly use more general formulas for $P$, $\varepsilon$ and $\mathcal{E}$.   However, in this case, it is unlikely that exact answers will be found.   In principle, it seems likely that the analogy to effective Newtonian mechanics in a conservative potential will survive, although the manipulations will become quite cumbersome.

Finally, we note that although we have only been discussing solutions with static dynamics, since our hydrodynamic equations enjoy Galilean invariance, we can easily generate solitons from any kink solution by simply moving to a new reference frame.

\subsection{Dynamics in a Magnetic Field}
In this subsection, we will consider how the competition between parity-violating Hall viscosity and magnetic forces in a charged parity-violating fluid can lead to new phenomena.  The key point of this analysis is that this combination of initial conditions with both $v_x$ and $v_y$ non-vanishing would be unstable in the parity-symmetric equations.   However, the Hall viscosity stabilizes the dynamics of the perturbations.

If we add a magnetic field $B$ to dissipationless flows, new interesting phenomena emerge.    Let us begin by looking for time-independent flows.   A simple solution to the equations of motion with no pressure gradients corresponds to \begin{equation}
v_y = V\cos(k^*x)   \label{vyb}
\end{equation}where \begin{equation}
k^* = \sqrt{\frac{\omega_{\mathrm{c}}}{\tilde{\nu}}}
\end{equation}and $n=n_0$, $v_x=0$.    Although this is also a solution to the dissipationless parity-symmetric fluid equations, here the interpretation is quite different.   Unlike in the dissipationless parity-symmetric case, the choice of $k^*$ is not arbitrary.    Therefore, this should not be thought of as a ``frozen" shear mode, but rather as a mode where ``two competing magnetic fields" cancel.   One magnetic field is the physical magnetic field $B$, while the other is an ``effective magnetic field" coming from the Hall viscosity, of strength $\tilde{\nu}k^2$.  Evidently, these fields can only cancel when $k=k^*$.    This gives us insight that one can think of Hall viscosity as a sort of wavelength-dependent magnetic field, which is helpful for understanding some of its curious properties.

We can also turn on a non-vanishing $v_x$, in addition to $v_y$ above:\begin{equation}
v_x = aV \cos(k^*y)  \label{vxb}
\end{equation}where $a$ is a dimensionless constant.   In the limit where $V\ll \sqrt{\tilde{\nu}\omega_{\mathrm{c}}}$ it is easy to check that the convective term in the momentum conservation equations is a perturbation compared to the parity-violating terms, and so approximately, this solution is also a static solution to the fluid equations.

In this approximation, the trajectories of tracer particles which obey $\dot{x}_i = v_i$ can easily be found to be \begin{equation}
a\sin(k^*x) - C = \sin(k^*y),
\end{equation} where $C$ is an integration constant.     In the symmetric case $a=1$, these trajectories split into two families, one which corresponds to sine-like waves following a line $y=n\pi/k^*$ ($n$ an integer), and another following the line $x=n\pi/k^*$, with unstable trajectories along straight lines when $C=0$.    The observation of tracers following these sine-like trajectories is evidence of the ``lattice-like" velocity structure that is stabilized in the parity-violating fluid, that may be realizable in experiments -- particularly with classical liquid crystals.

Now, let us study the perturbative correction to the fluid dynamics due to the convective term.   For simplicity, let us suppose that the velocity fields of Eqs. (\ref{vyb}) and (\ref{vxb}) are our initial conditions, with $a=1$.   We then expand $n=n_0+\delta n$, $v_x = v_x^0 + \delta v_x$, and $v_y = v_y^0+\delta v_y$.   As the perturbative equations are linear, we can also solve them by looking at one Fourier mode at a time.   As the convective term sources the perturbations, there are four Fourier modes excited at $(\pm k^*, \pm k^*)$.   Let us allow $\sigma_{x,y} \equiv \pm 1$, and study the $(\sigma_x k^*, \sigma_y k^*)$ mode.   The perturbative equations are \begin{subequations}\begin{align}
\partial_t \delta n + \mathrm{i} n_0 k^* (\sigma_x \delta v_x + \sigma_y \delta v_y) &= 0, \\
\partial_t \delta v_x + \frac{k^* V^2}{4} \sigma_y  &= -\mathrm{i}\sigma_x\frac{k^* c^2}{n_0} \delta n - \omega_{\mathrm{c}} \delta v_y , \\
\partial_t \delta v_y + \frac{k^* V^2}{4} \sigma_x  &= -\mathrm{i}\sigma_y\frac{k^* c^2}{n_0} \delta n + \omega_{\mathrm{c}} \delta v_x .
\end{align}\end{subequations}Write $\delta v_i = \delta v_i^\prime + \delta v_i^0$, where $\delta v_i^0$ is non-dynamical in time and given by \begin{subequations}\begin{align}
\delta v_y^0 &= \frac{k^* V^2}{\omega_{\mathrm{c}}} \sigma_x,\\
\delta v_y^0 &= -\frac{k^* V^2}{\omega_{\mathrm{c}}} \sigma_y,
\end{align}\end{subequations}so that the perturbation equations for $\delta v_x$, $\delta v_y$ and $\delta n$ are strictly linear.   The resulting dynamics will consist of velocity perturbations of size $k^*V^2/\omega_{\mathrm{c}}$, including a frozen $\omega=0$ mode and two dynamical modes at $\omega = \pm \sqrt{\omega_{\mathrm{c}}^2 + c^2 k^{*2}}$.

\section{Rayleigh-B\'enard Convective Instability}\label{sec6}
Our last example of a parity-violating flow is the Rayleigh-B\'enard convective instability in the Boussinesq approximation \cite{ chandrasekhar, fetter}.   Rayleigh-B\'enard convection is a classical instability of fluids that is readily observable in experiments, and so it is natural to question whether parity-violation can alter the nature of the instability.   In this section we will describe how parity-violation can either suppress or enhance the instability.   Our main evidence is provided by numerical solutions of the eigenvalue equations determining stability, though we will present some heuristic arguments for our observations as well.

The starting point of this problem is to consider a nearly incompressible fluid at rest, placed between two plates located at $y=\pm h/2$.   We assume that these plates are rigid, and that the plate at $y=-h/2$ is held fixed at temperature $T=T_0+\Delta T$, and the plate at $y=h/2$ has $T=T_0$;  we take $\Delta T >0$.   By nearly incompressible, we mean that we take \begin{equation}
n=n_0 (1-\alpha (T-T_0)),
\end{equation}  where $T_0$ is some reference temperature.   We take $\alpha \Delta T \ll 1$, so that essentially this temperature correction is negligible in most terms in the equations.   There is a solution to the equations of motion where the velocity is constant, with no electromagnetic fields and a temperature profile \begin{equation}
T - T_0 = \Delta T \left(\frac{1}{2}-\frac{y}{h}\right).
\end{equation}We assume that there is a gravitational field of strength $g$ that opposes the temperature gradient.    The pressure profile is given by noting that \begin{equation}
\partial_y P = - mn_0 \left(1-\alpha \Delta T \left(\frac{1}{2}-\frac{y}{h}\right)\right) g,
\end{equation}which can be integrated straightforwardly.   The basic idea is that this flow can be unstable:  a perturbation which induces non-trivial velocity flows can enhance the transfer of energy between the two plates beyond the limit set by thermal diffusion.

Now, let us ask for the stability of this flow by perturbing it.   The Boussinesq approximation is that the \emph{only} place where $\alpha$ must not be neglected in the fluctuation equations is in the fluctuations in gravitational forces due to temperature (and the resulting density) fluctuations.   Linearizing the parity-violating hydrodynamics around this background solution with the Boussinesq approximation, and transforming into Fourier modes $\mathrm{e}^{\mathrm{i}kx-\mathrm{i}\omega t}$, we find \begin{subequations}\begin{align}
\mathrm{i}k \delta v_x + \partial_y \delta v_y &= 0, \\
- \mathcal{C}_T \left(\mathrm{i}\omega \delta T + \frac{\Delta T}{h}\delta v_y\right) - \mathrm{i}\omega \frac{mf(T_0)}{T_0}\frac{\Delta T}{h} \delta v_x + \kappa \left(k^2-\partial_y^2\right) \delta T &= 0, \label{eq89b} \\
-\mathrm{i}\omega \delta v_x + \mathrm{i}k\frac{\delta P}{mn_0} + \nu \left(k^2-\partial_y^2\right) \delta v_x  + \tilde{\nu} \left(k^2-\partial_y^2\right) \delta v_y &= 0, \\
-\mathrm{i}\omega \delta v_y + \partial_y\frac{\delta P}{mn_0}  - \alpha g \delta T  + \nu \left(k^2-\partial_y^2\right) \delta v_y  - \tilde{\nu} \left(k^2-\partial_y^2\right) \delta v_x &= 0.
\end{align}\end{subequations}As the plates are rigid with fixed temperature, we impose $\delta v_x = \delta v_y = \delta T = 0$ at both $y=0, h$.   We have also neglected many terms in the momentum equation proportional to $\alpha$, and a term proportional to $\mathcal{C}_n\alpha$, as is common in a first treatment \cite{chandrasekhar, fetter}.    It is convenient to non-dimensionalize the problem.   By rescaling \begin{subequations}\begin{align}
x_i &\rightarrow hx_i, \\
t &\rightarrow \frac{h^2}{\nu} t, \\
T &\rightarrow \frac{\nu \mathcal{C}_T \Delta T}{\kappa} T,
\end{align}\end{subequations}we can express the dynamics in terms of universal, dimensionless parameters.   It is also helpful to re-express things in terms of a stream function: $\delta v_y =- \mathrm{i}k \delta \psi$, $\delta v_x = \partial_y \delta \psi$.  Putting this all together, we find that (remember -- everything is now in dimensionless units): \begin{subequations}\label{conveig}\begin{align}
-\mathrm{i}\omega \left(\partial_y^2 - k^2\right) \delta \psi - \left(\partial_y^2 - k^2\right)^2 \delta \psi + \mathrm{i}kR\delta T &= 0, \label{eq92a} \\
-\mathrm{i}\omega Q \delta T + \mathrm{i}k \delta \psi + \left(k^2-\partial_y^2\right)\delta T - \mathrm{i}\omega G \partial_y\delta \psi&= 0, \label{eq92b}
\end{align}\end{subequations} where we have defined three dimensionless numbers: \begin{subequations}\begin{align}
R &\equiv \frac{\alpha g\mathcal{C}_Th^3}{\nu\kappa}, \\
Q &\equiv \frac{\nu\mathcal{C}_T}{\kappa}, \\
G &\equiv  \frac{mf(T_0)\nu}{\mathcal{C}_TT_0h^2}.
\end{align}\end{subequations}Note that as this flow is treated as incompressible, as we expect, the Hall viscosity does not play any role.  Eq. (\ref{conveig}), together with our boundary conditions, is a generalized eigenvalue problem that can efficiently be solved with the same numerical methods discussed previously.

As with flow down an inclined plane, it is easy to mock up a gravitational field in a charged fluid with an electric field:\footnote{This is known to work as well in liquid crystals \cite{tavener}.} \begin{equation}
E_y = -\frac{mg}{q}.
\end{equation}We will still, for simplicity, express answers in terms of this effective $g$.   However, quick inspection of Eq. (\ref{eqencons}), conservation of energy, reveals that some new terms will also contribute.   Eq. (\ref{eq89b}) becomes \begin{equation}
- \mathcal{C}_T \left(\mathrm{i}\omega \delta T + \frac{\Delta T}{h}\delta v_y\right) - \mathrm{i}\omega \frac{mf(T_0)}{T_0}\frac{\Delta T}{h} \delta v_x + \kappa \left(k^2-\partial_y^2\right) \delta T - \mathrm{i}k mg \left(f^\prime(T_0)+\frac{f(T_0)}{T_0}\right)\delta T = 0.
\end{equation} Eq. (\ref{eq92b}) then becomes \begin{equation}
-\mathrm{i}\omega Q \delta T + \mathrm{i}k \delta \psi + \left(k^2-\partial_y^2\right)\delta T - \mathrm{i}\omega G \partial_y\delta \psi - \mathrm{i}k F\delta T= 0,
\end{equation}where we have defined a new dimensionless parameter \begin{equation}
F \equiv \left(f^\prime(T_0) + \frac{f(T_0)}{T_0}\right) \frac{mg\nu \mathcal{C}_T\Delta T}{\kappa h}.
\end{equation}
$F=G=0$ corresponds to the solution of the classic parity symmetric problem; otherwise, we are solving a parity-violating counterpart.

\begin{figure}[h!]
\centering
\includegraphics{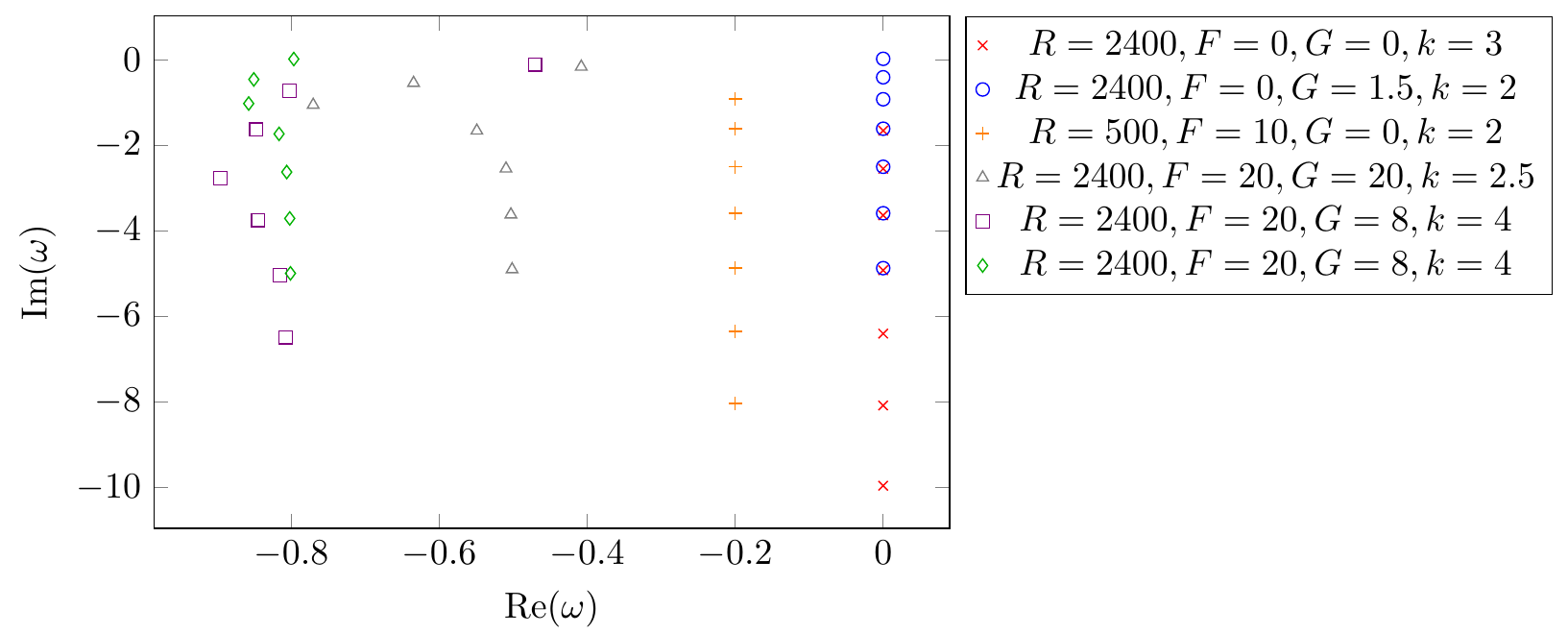}
\caption{The 7 most unstable modes in the Rayleigh-B\'enard problem for select values of $R$, $G$, $F$ and $k$.  So that all data fits in the same plot, we have had to choose the same value $Q=100$ for all data, but we have observed that the qualitative types of behavior depicted above occur for other values of $Q$.}
\label{conveigfig}
\end{figure}

\begin{figure}[h!]
\centering
\includegraphics{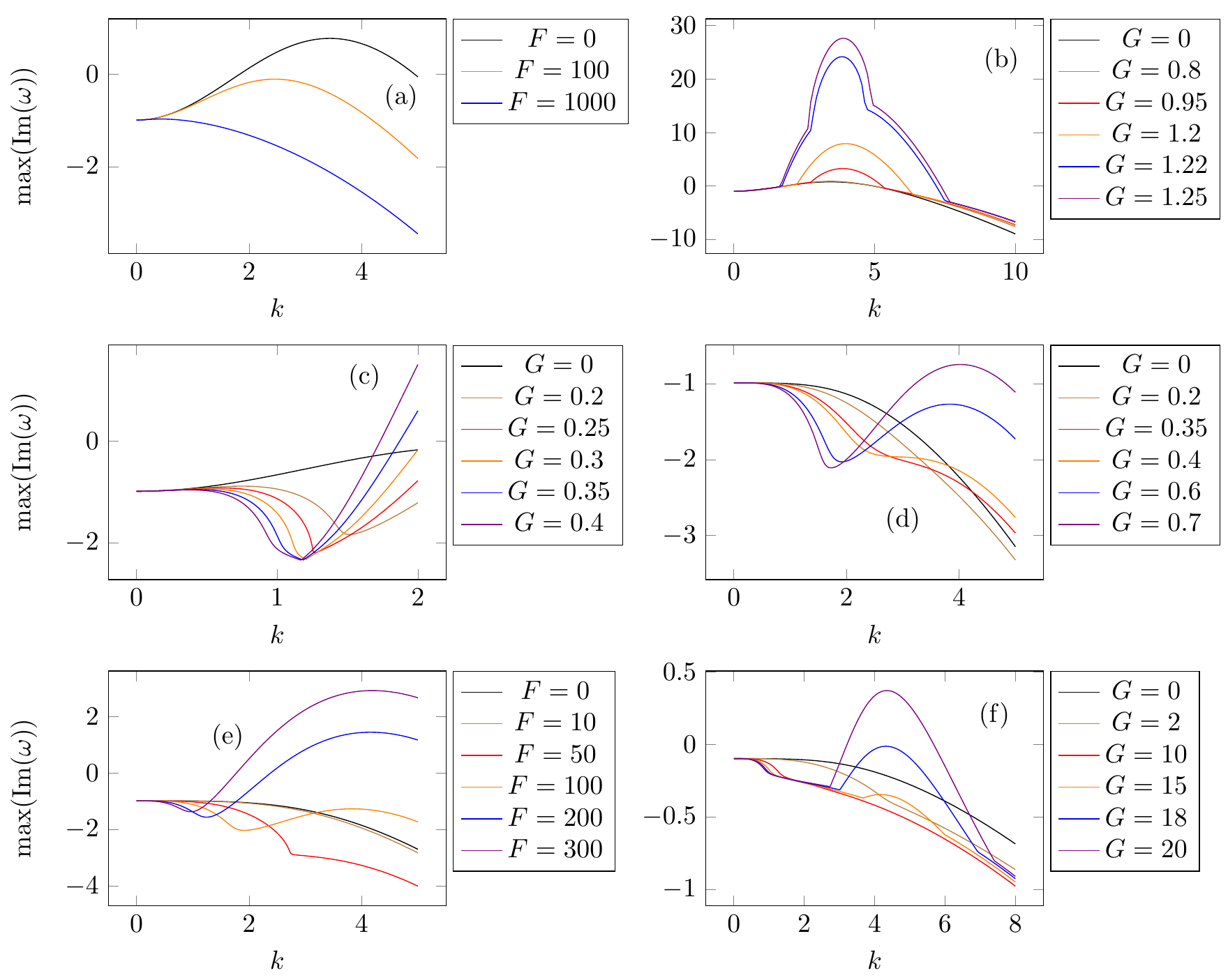}
\caption{Plots of the most unstable mode in the Rayleigh-B\'enard problem for a variety of parameters.   Here are the parameters which are held fixed in each plot:  (a): $R=2400$, $Q=10$, $G=0$, (b): $R=2400$, $Q=10$, $F=0$,  (c): $R=2400$, $Q=10$, $F=100$, (d): $R=500$, $Q=10$, $F=100$, (e): $R=500$, $Q=10$, $G=0.6$, (f): $R=500$, $Q=100$, $F=100$. }
\label{figrb}
\end{figure}

We now present results from a numerical calculation the eigenfrequencies $\omega$.   Here we used a smaller number $N=28$ of Chebyshev polynomials.   Let us begin by studying when $G=0$, and $F\ne 0$.   Firstly, we expect the leading order effect at small $F$ is to see chiral corrections to the dispersion relation $\omega(k)$:  indeed, we find this is borne out in Figure \ref{conveigfig}.   As we expect, the size of $\mathrm{Re}(\omega) \sim F$, and the corrections to $\mathrm{Im}(\omega)\sim F^2$, for $F\lesssim 1$.    Secondly, we find that beyond leading order in perturbation theory, $F$ always a suppression of the onset of instability, as we show in Figure \ref{figrb}a.   A heuristic explanation for this is that the chiral propagation of fluctuations can help to ``carry away" a fluctuation before it has time to develop into an instability.   Finally, note that changing $k\rightarrow -k$ and $F\rightarrow -F$ leaves $\omega$ invariant, explaining why we have only studied $F\ge 0$.

Before moving on, let us note a curious feature of the dispersion relations at finite $F$.   Unless $G$ is reasonably large, we find that $\mathrm{Re}(\omega)$ is approximately constant.   In fact, this constant is approximately given by a universal formula: \begin{equation}
\mathrm{Re}(\omega) =  -\frac{F}{Q}k.   \label{eqfqk}
\end{equation}Here is a semi-quantitative reason why this should be the case.   Let us consider the problem with $G=0$, and choose the boundary conditions so that $\partial_y v_x =0$ at $y=0,h$.   This problem is somewhat unphysical, but in this case one can easily see from the perturbation equations that choosing both $\delta\psi$ and $\delta T$ to be proportional to $\sin (n\pi y)$ satisfies all boundary conditions, and is also consistent with the perturbation equations.      $\omega$ can also be exactly found.   There are always a pair of normal modes associated with a given choice of $n$ and $k$, and: \begin{align}
\omega  &= -\frac{kF-\mathrm{i}(1+Q)\left(k^2+(n\pi)^2\right)}{2Q} \notag \\
&\pm \dfrac{\mathrm{i}}{2Q}  \sqrt{\left((1-Q)^2-k^2F^2\right)\left(k^2+(n\pi)^2\right)^2 + 2(Q-1)\mathrm{i}kF\left(k^2+(n\pi)^2\right)+ \dfrac{4RQk^2}{k^2+(n\pi)^2}}
\end{align}Expanding this equation to first order in $F$, in the limit of $1\ll Q$ and $F \ll Q$, we approximately recover Eq. (\ref{eqfqk}).   Also, to higher orders in $F$, it is straightforward from this equation to plot the resulting $\omega$ numerically and see that we also find that $\mathrm{Im}(\omega)$ (for the more unstable ($+$) branch of solutions) decreases as $F$ increases.

Next, let us consider the case where $F=0$, but $G\ne 0$, as shown in Figure \ref{figrb}b.    Again, we only have to consider $G\ge 0$, as changing the sign of both $y$ and $G$ leaves the problem invariant.   Although $F$ tends to suppress instability, we find that $G$ (alone) enhances this instability. Here, we find a very curious effect:  there are \emph{discontinuous derivatives} in $\max(\mathrm{Im}(\omega(k)))$.   We have resolved these points to high precision numerically and this effect does not go away.   Although we do not see this effect for small $G$, at larger $G$ we see the onset of the new ``branch", and at even larger $G$ we see a second ``branch" emerge.   It is unclear whether three is the maximal number of such ``branches".  We propose an explanation for this effect in Appendix \ref{appb}.   Also note, as shown in Figure \ref{conveigfig}, that we have not found any real part to $\omega$, even when $G\ne 0$.

Next, let us consider allowing both $F\ne 0$ and $G\ne 0$.   The remaining four panels of Figure \ref{figrb} are dedicated to this case.   We do not have a universal understanding of the behavior of $\omega$ in this four-parameter space, so let us simply elucidate some features that we found interesting.   In some cases, we found that the discontinuous derivatives at large values of $G$ persisted at finite $F$, and in other cases they did not.   Also, whereas when $F=0$, $G$ always enhanced the instability, in some cases we find that $G$ can enhance the instability at some values of $k$, and suppress it at others.

\begin{figure}[h!]
\centering
\includegraphics[width=7in]{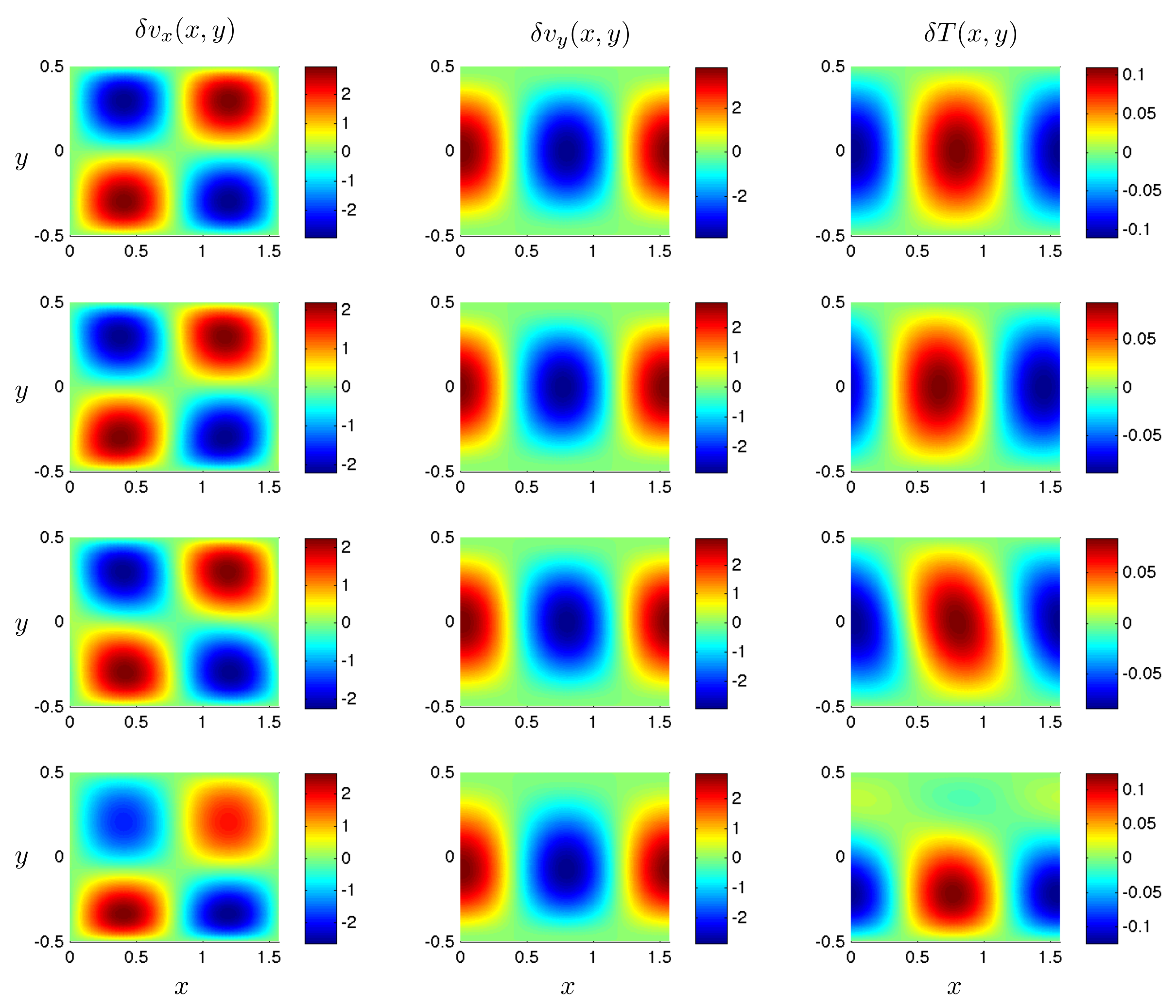}
\caption{Plots of the most unstable eigenvector in the Rayleigh-B\'enard problem for fixed $k=4$, $R=2400$, $Q=10$.   Rows from top to bottom:  $F=G=0$;  $F=50$, $G=0$;  $F=0$, $G=0.4$;  $F=20$, $G=0.2$.}
\label{eigenvectorfig}
\end{figure}

Finally, in Figure \ref{eigenvectorfig} we show two dimensional plots of the most unstable eigenvectors for fixed $k$, $R$, and $Q$, and various $F$ and $G$.    We look and $F$ and $G$ small enough so that the eigenvectors are similar to the parity-symmetric case (and that the most unstable mode is the same), but large enough so that parity-violating effects are easily visible.   It is evident in each case that parity violation has distorted the unstable modes by breaking their invariance under parity.   The case where $F\ne 0$ and $G=0$ shows a phase shift of the thermal unstable mode, relative to the velocities;  when $F=0$ but $G\ne 0$, the fluctuations are sheared relative to the parity  symmetric case.   The case $F\ne G\ne 0$ in particular displays a dramatic enhancement of the unstable modes for $y<0$.   Let us not forget that in addition to this, these modes also have $\mathrm{Re}(\omega)\ne 0$ when $F\ne 0$, implying that the instability propagates along the channel.   This is a dramatic difference to the parity symmetric case, where the initial instability does not propagate in the $x$-direction.  In principle, these would be the modes observed at early times during the unstable dynamics of the fluid, though at late times the convective patterns may look quite different \cite{bodenschatz}.   Nonetheless, it would not be surprising if parity was broken in the late time dynamics as well, providing a dramatic dynamical method of observing the broken symmetry.

\section{Conclusion}
Our main purpose in this paper has been to explore the phenomenological consequences of new parity-violating terms in the gradient expansion of hydrodynamics in 2+1 dimensions.   Although the Hall viscosity is notoriously challenging to measure for incompressible fluids, we have suggested a variety of mechanisms by which the Hall viscosity contributes to qualitatively new dynamics.

There are many other possible choices of a hydrodynamic gradient expansion, both in 2+1 dimensions and in higher dimensions, whose phenomenology has not been fully explored.   One obvious direction is to extend this formalism to 3+1 dimensions, where new manifestations of parity-violation occur, and new phenomena should result.   There are two examples of such phenomena that has already been explored in the literature: the chiral magnetic effect \cite{cme2, cme1} and the chiral vortical effect \cite{cve1, cve2}, though these are both static effects.  Given the possibility of exotic new types of hydrodynamic flows, extending the ideas presented in this work to these other theories is an interesting direction for future work.

One subtlety in the above formalism the role played by the $f(T)$ term.   We have seen that this term is responsible for much of the exotic dynamics of parity-violating fluids.   It does seem rather curious that there is still a finite charge current in a temperature gradient, independent of $n$.   This suggests that quantum, relativistic microscopic dynamics may be required for the term to make sense: such a current could occur because transport is governed by particles and holes\footnote{Note that, counterintuitively from a non-relativistic perspective, particles and holes (anti-particles) contribute to $n$ with opposite signs.} moving in opposite directions, and therefore for a classical liquid crystal film, perhaps $f=0$.  See \cite{banerjee2} for further discussions on constraints on parity violating transport coefficients.   A deeper understanding of the microscopic origins of each parity-violating term in the above gradient expansion (even at the relativistic level) may shed light on such concerns.

Another subtlety is that for much of this paper, we have assumed ``standard" boundary conditions of fluid mechanics -- e.g., no-slip (Dirichlet) boundary conditions on velocity fields, or stress-free boundary conditions.   These boundary conditions may change for some parity-violating fluids; if so, the solutions we have written down above must be altered.   This is a worthwhile direction for further investigation.

Given that the study of the classical Navier-Stokes equations is an entire discipline of physics, engineering and mathematics, it seems natural to ask phenomenological questions about parity-violating hydrodynamics purely for curiosity's sake.    However, it may also be possible to find powerful signatures of parity-violation through hydrodynamic phenomena which are  unique to parity-violating fluids.  Along these lines, it is worth noting that many of the particular flows that we constructed in this paper are quite easy to construct for parity-symmetric classical fluids such as water or air in the laboratory (and are usually done in three spatial dimensions), and it may be possible using present day technology to construct some flows using two-dimensional films of liquid crystals: for example, Rayleigh-B\'enard convection in liquid crystals \cite{leif, tavener}.   However, these flows would likely be very challenging to construct in a laboratory for a strongly-coupled quantum fluid.    Therefore, we hope that the simple examples in this paper, which can be solved either analytically or with very short numerical codes, provide toy models where insight into more complicated solutions to the equations of parity-violating hydrodynamics -- which may be more easily realized in experiments -- is possible.

\section*{Acknowledgements}
\addcontentsline{toc}{section}{Acknowledgements}
We would like to thank Alexander Abanov, Matthias Kaminski, Xiao-Liang Qi, Koenraad Schalm, Paul Wiegmann, and the referees for helpful comments.

A.L. is supported by the Smith Family Graduate Science and Engineering Fellowship.   The research of P.S. was supported by a Marie Curie International Outgoing Fellowship, grant number PIOF-GA-2011-300528.

\begin{appendix}
\titleformat{\section}
  {\gdef\sectionlabel{}
   \Large\bfseries\scshape}
  {\gdef\sectionlabel{\thesection. }}{0pt}
  {\begin{tikzpicture}[remember picture,overlay]
	\draw (-0.2, 0) node[right] {\textsf{Appendix \sectionlabel#1}};
	\draw[thick] (0, -0.4) -- (\textwidth, -0.4);
       \end{tikzpicture}
  }
\titlespacing*{\section}{0pt}{15pt}{20pt}

\section{Perturbative Approach for Normal Modes} \label{appa}
Because the parity-violating contributions to the hydrodynamic equations generically couple together all fields in a non-trivial way, it is generally impossible to find analytical solutions for the normal modes of a set-up.   In this appendix, we present a way of calculating perturbatively the normal modes when the parity-violating terms are treated as small, that we have found especially helpful for systems which break chiral symmetry.   To do so, we use an analogy to perturbation theory in quantum mechanics \cite{Landau1977}.\footnote{This approach was used  in \cite{lucasunpublished}, in the context of waveguides with a normal insulator-topological insulator boundary, based off of the modified electrodynamics of \cite{qi}.}

Suppose for simplicity that the equations of motion are translation invariant in the $x$ and $t$ directions, so that we may study normal modes of the form $\psi(y) \mathrm{e}^{\mathrm{i}kx-\mathrm{i}\omega t}$.   In this case, the normal mode equations reduce to finding the kernel of a  differential operator $\mathcal{L}_0$ in a single variable: \begin{equation}
\mathcal{L}_0\psi = 0.
\end{equation}As a simple example of this, consider the operator $\mathcal{L}_{\mathrm{inc}}$ which corresponds to the incompressible dynamics of the stream function: \begin{equation}
\mathcal{L}_{\mathrm{inc}}(\omega) = \left(k^2- \partial_y^2\right)\left(\nu k^2 - \nu \partial_y^2 - \mathrm{i}\omega \right) .
\end{equation}This kernel is not a trivial product because we impose 4 boundary conditions on the stream function $\psi$.    Also, in general, this kernel is trivial -- only at special values of $\omega$, associated to a normal mode, will this kernel be non-trivial.    Assume for simplicity that the normal mode frequencies are separated by some finite frequency shifts, so that the kernel of $\mathcal{L}_{\mathrm{inc}}(\omega)$ always has dimension of either 0 or 1.   And although it may not appear so, this operator is Hermitian -- $\omega$ is purely imaginary for normal modes.

Now, suppose that we shift our differential operator to $\mathcal{L}_0(\omega) + \epsilon \mathcal{L}_1$, where $\epsilon$ is an infintesimal number.   The operator $\mathcal{L}_1$ may be a completely arbitrary function of $\partial_y$, $k$ and $\omega$ -- even one that is not a Hermitian operator, since we will only be interested in the lowest order effects.  The normal mode frequencies will shift to $\omega_0 + \epsilon \omega_1$.   Now, let us denote abstractly the $n^{\mathrm{th}}$ normal mode of $\mathcal{L}_0$ by $|n_0\rangle$, just as in quantum mechanics.   Denoting \begin{equation}
\langle m|n\rangle \equiv \int \mathrm{d}y\; \overline{\psi}_m \psi_n \label{eq88}
\end{equation}and $\omega_{n,0}$ as the frequency of this normal mode ($\mathcal{L}_0(\omega_{n,0})|n_0\rangle = 0$), so long as $\mathcal{L}_0$ is Hermitian, eigenfunction orthogonality implies $\langle m_0|n_0\rangle = \delta_{mn}$.    Thus, we know from our analogy with quantum mechanical perturbation theory that the first order correction to the eigenvalue of $\mathcal{L}_0+ \epsilon \mathcal{L}_1$ is simply given by $\langle n_0 | \mathcal{L}_0+\epsilon \mathcal{L}_1 |n_0\rangle$.   Denoting $\mathcal{L}_0^\prime \equiv \partial_\omega \mathcal{L}_0$, we can straightforwardly conclude that \begin{equation}
\omega_{n,1} = -\frac{\langle n_0|\mathcal{L}_1|n_0\rangle}{\langle n_0 | \mathcal{L}_0^\prime |n_0\rangle}.  \label{eq89}
\end{equation}

We have used this perturbation theory to study the small chiral corrections to the normal modes of our waveguide.  Let us begin with Eq. (\ref{wgeq}), which describes the waveguide with a parity-violating interface.   Here our perturbative parameter will be $\tilde{\nu}$, so we begin by setting it to 0.  In this case, reflection symmetry about $y=0$ splits the normal modes into even and odd modes.  For simplicity, let us start by focusing on the even modes, which are given by \begin{equation}
\psi = \frac{\cosh(ky/2)}{\cosh(kh/2)} - \frac{\cosh(qy/2)}{\cosh(qh/2)}.
\end{equation}with $k\ne q$.\footnote{One can easily show that there are too many boundary conditions for a ``special" normal mode with $k=q$ and $\omega=0$ to exist.   This is also true for the odd modes.}  The boundary conditions that $\partial_y\psi$ vanish at the boundary correspond to a normal mode equation: \begin{equation}
\frac{kh}{2}\tanh\frac{kh}{2} = \frac{qh}{2}\tanh\frac{qh}{2},\label{eq91}
\end{equation}The function $x\tanh x$ is monotonically increasing, so evidently $\mathrm{Im}(q) \ne 0$.   In fact, we have found that the solutions to Eq. (\ref{eq91}) have $\mathrm{Re}(q)=0$, as well as $\mathrm{Re}(\omega) =0$.  Evidently $\mathcal{L}_0$ is still Hermitian.  After a straightforward calculation, one finds that \begin{equation}
\omega_1 =  \frac{2k^3\tilde{\nu}\nu}{-\mathrm{i}\omega} \left(\mathrm{sech}\frac{kh}{2} - \mathrm{sech}\frac{qh}{2}\right)^2 \left(\frac{2}{k-q}\sinh \frac{kh-qh}{2} + \frac{2}{k+q}\sinh \frac{kh+qh}{2} -\frac{1}{q}\sinh\frac{qh}{2}-\frac{h}{2}\right)^{-1}.
\end{equation}This expression tells us that $\omega_1$ is real; for $\tilde{\nu}$ and $k$ positive, $\omega_1<0$.   The odd modes are given by \begin{equation}
\psi = \frac{\sinh(ky/2)}{\sinh(kh/2)} - \frac{\sinh(qy/2)}{\sinh(qh/2)}.
\end{equation}Since $\psi(0)=0$ for the odd modes, $\omega_1=0$ for these modes, so the corrections to $\omega$ are $\mathrm{O}(\tilde{\nu}^2)$.

We have checked this analytic result against our numerics.   In the perturbative limit where $a,k \lesssim 1$, we find very consistent agreement that $\mathrm{Re}(\omega_1) \sim ak^3$, as well as $\mathrm{Im}(\omega_1) \sim a^2$.    However, the complicated prefactor we have predicted is typically overshot by a factor of about 5\%.   We suspect that this is due to subtleties resolving the $\delta$ function.


\section{A Toy Model of Generalized Eigenvalue Problems} \label{appb}
We have seen exotic behavior in Sec. \ref{sec6} where $\max(\mathrm{Im}(\omega))$ jumps suddenly as we tune the value of $G$.   Here we give a simple, heuristic argument for this effect.

Let us consider a very simple toy model of a generalized eigenvalue problem involving differential operators.   Consider the function $\psi(y)$ defined on the line $y\in [0,1]$, with the boundary conditions $y(0)=y(1)=0$.   Our generalized eigenvalue problem is \begin{equation}
-\partial_y^2 \psi = \lambda \left(1+\mu \partial_y\right)\psi
\end{equation}for the generalized eigenvalue $\lambda$.   We assume that $\mu$ is real.   This can be exactly solved by noting that $\psi = A\mathrm{e}^{\mathrm{i}q_+y} + B\mathrm{e}^{\mathrm{i}q_-y}$ where \begin{equation}
q_\pm = \frac{\mathrm{i}\mu\lambda \pm \sqrt{4\lambda-\lambda^2\mu^2}}{2}.
\end{equation}The eigenfunctions are \emph{real}, and are given by \begin{equation}
\psi = \mathrm{e}^{-\mu \lambda y/2} \sin (n\pi y), \label{eq107}
\end{equation}where \begin{equation}
2\pi n = \sqrt{4\lambda - \lambda^2\mu^2}.
\end{equation}The solutions to this are given by \begin{equation}
\lambda = \frac{2\left(1\pm \sqrt{1-(n\pi\mu)^2}\right)}{\mu^2}.
\end{equation}Note that because $\lambda$ shows up in Eq. (\ref{eq107}), these correspond to two distinct eigenfunctions.

At $\lambda=0$, ``one half" of the eigenvalues are at $\infty$, and we thus ignore them.   At small $\mu$, the smallest value of $\lambda$ is given by $2\mu^{-2} \left(1-\sqrt{1-(\pi \mu)^2}\right)$;  as $\mu\rightarrow 0$, this reduces to approximately $\pi^2$ as we expect.   When $\mu \ge 1/\pi$, however, we see that $\min(\mathrm{Re}(\lambda)) = 2\mu^{-2}$.   Thus we see a discontinuous jump in the first derivative of $\mathrm{Re}(\lambda)$ and $\mathrm{Im}(\lambda)$.   This gives a heuristic explanation of why such a jump could be observed in the more complicated parity-violating Rayleigh-B\'enard generalized eigenvalue problem.

There is one important difference between this toy model and the Rayleigh-B\'enard problem.   Here, at the point where the solution qualitatively changes, the eigenvalues suddenly pick up imaginary parts.    We do not find an analogous effect for the Rayleigh-B\'enard problem -- to many orders of magnitude, we find $\mathrm{Re}(\omega)\approx 0$.
\end{appendix}

\bibliographystyle{unsrt}
\addcontentsline{toc}{section}{References}
\bibliography{paritybib}

\end{document}